\documentclass[aps,reprint,prd,amsmath,amssymb,nofootinbib,showpacs]{revtex4-1}
\usepackage[english]{babel}
\usepackage{graphicx}
\usepackage{bm}
\usepackage{epsf}

\usepackage{color}

\usepackage[normalem]{ulem}  
\newcommand{\old}[1]{}

\begin{document}

\title{Transport coefficients of leptons in superconducting neutron star cores}
\author{P.~S. Shternin}\email{pshternin@gmail.com}
\affiliation{Ioffe Insitute, 26 Politekhnicheskaya st., St. Petersburg, 194021, Russia}

\date{\today}


\begin{abstract}
I consider the thermal conductivity and shear viscosity of leptons (electrons and muons) in the nucleon
neutron star
cores where protons are in the superconducting state. I restrict the consideration to the case of not too high temperatures $T\lesssim 0.35T_{\mathrm{c}p}$, where $T_{\mathrm{c}p}$ is the critical temperature of the proton pairing. In this case, lepton collisions with protons can be neglected. Charged lepton collision frequencies are mainly determined by the transverse plasmon exchange and are mediated by the character of the transverse plasma screening. In our previous works [Shternin \& Yakovlev, Phys. Rev. D {\bf 75} 103004 (2007); {\bf 78} 063006 (2008)] the superconducting proton contribution to the transverse screening was considered in the Pippard limit $\Delta \ll \hbar q v_{\mathrm{F}p}$, where $\Delta$ is the proton pairing gap, $v_{\mathrm{F}p}$ is the proton Fermi velocity, and $\hbar q$ is the typical transferred momentum in collisions. However, for large critical temperatures (large $\Delta$) and relatively small densities (small $q$) the Pippard limit may become invalid. In the present study I show that this is indeed the case and that the older calculations severely underestimated the screening in a certain range of the parameters appropriate to the neutron star cores. As a consequence,  the kinetic coefficients at $T\ll T_{\mathrm{c}p}$ are found to be smaller than in previous calculations.
\end{abstract}
\maketitle

\section{Introduction}\label{sec:intro}

Neutron stars (NSs) are the most compact stars known in the
Universe comprising about 1.5 solar masses in a $\sim 12$~km
radius sphere. In their interiors, NSs  contain superdense matter
of largely unknown composition  \cite{HPY2007Book}. Their
astrophysical manifestations are numerous, delivering signals in
all bands of the electromagnetic spectra \cite{Kaspi2010PNAS}.
Moreover, gravitational waves from a binary NS merger  were
detected recently \cite{AbbotPhysRevLett}. Understanding NSs
requires modelling of various processes in their interiors.
Important ingredients for this modelling are the transport
coefficients of the superdense matter
\cite{SchmittShternin2017arXiv,PotekhinSSR2015}.

In the present paper I discuss the thermal conductivity $\kappa$
and shear viscosity $\eta$ in NS cores of the simplest composition
containing mainly neutrons (n) with admixture of protons (p),
electrons (e), and muons ($\mu$). Electrons and muons form
relativistic degenerate almost ideal Fermi gases, while baryons
(neutrons and protons) form non-ideal strongly-interacting Fermi
liquid \cite{HPY2007Book}. Transport coefficients are governed by
the particle collisions. Leptons collide with themselves and with
charged protons due to electromagnetic interaction, while
collisions between baryons are mediated mainly by the strong
interaction. To a good approximation, it is possible to consider
lepton and baryon subsystems separately \cite{FlowersItoh1979ApJ}.
For instance for the thermal conductivity one writes
$\kappa=\kappa_{e\mu}+\kappa_{np}$. In this case, when the lepton
part, $\kappa_{e\mu}$ (or $\eta_{e\mu}$), is calculated, protons
(or other charged baryons if present) are treated as passive
scatterers.

Currently adopted calculations of the lepton contribution to
transport coefficients of non-superfluid NS core matter were
performed in
Refs.~\cite{ShterninYakovlev2007PhRvD,ShterninYakovlev2008PhRvD,Shternin2008JETP}
with a proper account for the screening of electromagnetic
interaction following original ideas of
\citet{Heiselberg1992NuPhA} and \citet{Heiselberg:1993cr}.
Calculations of the nucleon part, $\kappa_{np}$ and $\eta_{np}$,
are more uncertain since one needs to rely on a certain many-body
theory of nuclear matter. Transport coefficients in the nucleon
sector are studied, for instance, in
Refs.~\cite{Shternin2013PhRvC,Kolomeitsev2015PhRvC,BenharValli2007PhRvL,Zhang2010PhRvC}
and more complete list of references can be found in the recent
review~\cite{SchmittShternin2017arXiv}.

Nuclear matter in NS cores can be in the superfluid (paired) state
due to an attractive component of the nuclear interaction
\cite{LombardoSchulze2001LNP,Page:2013hxa,Haskell2017arXiv,Sedrakian2018arXiv}.
Critical temperatures of the proton paring $T_{\mathrm{c}p}(n_B)$
and neutron pairing $T_{\mathrm{c}n}(n_B)$ depend on the baryon
number density $n_B$. Neutrons are believed to be paired in the
singlet $^1$S$_0$ state at low densities (low Fermi momenta). In
most models this type of the neutron pairing realizes in the NS
inner crust, where the gas of free unbound neutrons coexists with
the Coulomb lattice of ions and the degenerate electron gas. The
singlet neutron pairing ceases in the core, where the $^1$S$_0$
channel of the nuclear interaction becomes repulsive. Instead, the
neutron-neutron interaction becomes attractive in the triplet
$^3$P$_2$ channel leading to the anisotropic paired state in the
NS core. Proton number density is $\sim10$ times smaller than the
neutron one, therefore protons in the outer core are thought to be
paired in the $^1$S$_0$ channel. In the inner core, where the
proton number density increases, the $^1S_0$ proton pairing is
thought to disappear. Calculations of critical temperature
profiles for triplet neutron and singlet proton pairings in the NS
core are very model-dependent
\cite{LombardoSchulze2001LNP,Page:2013hxa,Gezerlis2014arXiv}.
Generally, the profiles $T_{\mathrm{c}p}(n_B)$ and
$T_{\mathrm{c}n}(n_B)$ are bell-like, reaching maximum at some
density within the core. The maximal critical temperature for
protons is thought to be in the range $10^9-10^{10}$~K, while for
triplet neutron superfluidity the corresponding values are found
to be  generally smaller, in the range of $10^8-5\times 10^9$~K.
For typical temperatures in the interiors of not too young NSs,
$T\sim 10^8$~K \cite{YakovlevPethick2004ARA}, protons in a large
part of the core are expected to be in the paired and hence
superconducting state.

Neutron superfluidity does not produce immediate effect on the
lepton contribution to the transport coefficients. In contrast,
the superfluidity of protons affects $\kappa_{e\mu}$ and
$\eta_{e\mu}$ in two aspects. The first one is the damping of the
lepton-proton collisions due to the reduction of the number of the
proton excitations. The lepton-proton scattering is damped roughly
by the exponential factor $\mathrm{exp}(-\Delta/T)$, where
$\Delta$ is the gap in the proton energy
spectrum.\footnote{Throughout the paper the natural unit system is
used, where $\hbar=c=k_B=1$.} The second effect comes from the
modification of the screening of the electromagnetic interactions
which affects collisions between all charged particles including
unpaired ones (leptons in the present case). Both these effects
were investigated in
Refs.~\cite{ShterninYakovlev2007PhRvD,ShterninYakovlev2008PhRvD}.
In these papers, the proton contribution to screening was taken in
the so-called Pippard limit, $q v_{\mathrm{F}p}\gg\Delta$, where
$v_{\mathrm{F}p}$ is the proton Fermi velocity and $q$ is the
momentum transfer in collisions, both of which increase with
density. In the present paper I show that this limit is
inapplicable for the wide range of conditions relevant for NS
cores, i.e. for not too high densities (small $q v_{\mathrm{F}p}$)
or for relatively high gap values (high $T_{\mathrm{c}p}$). The
opposite, London limit, $\Delta \gg q v_{\mathrm{F}p}$ can be
equally relevant for lepton scattering, and the transition between
two limiting cases occurs roughly at the transition between the
superconductors of the first and second kind.

The paper is organized as follows. In Sec.~\ref{sec:kin_form} the
general formalism needed to calculate transport coefficients of
npe$\mu$ matter of NS cores is briefly outlined and the results of
Refs.~\cite{ShterninYakovlev2007PhRvD,ShterninYakovlev2008PhRvD}
for a normal (non-superfluid) case are reviewed. In
Sec.~\ref{sec:prot_screen} the plasma screening properties in
presence  of the proton pairing are discussed and in
Sec.~\ref{sec:collfreq_lead}--\ref{sec:corr_Var} transport
coefficients in this case are calculated. The results are
summarized and discussed in Sec.~\ref{sec:discuss}. I conclude in
Sec.~\ref{sec:conclusions}.

The consideration in this study is limited to small temperatures,
$T\lesssim 0.35 T_{\mathrm{c}p}$ and the effects of magnetic
fields are not included.

\section{General expressions}\label{sec:kin_form}
Transport coefficients in NS cores can be calculated in the
framework of the transport theory of Fermi liquids
\cite{BaymPethick} adapted for multicomponent systems
\cite{FlowersItoh1979ApJ,Anderson1987,SchmittShternin2017arXiv}.
Below I closely follow
Refs.~\cite{ShterninYakovlev2007PhRvD,ShterninYakovlev2008PhRvD}
and omit the details.

Thermal conductivity $\kappa_c$ and shear viscosity $\eta_c$ of
particle species $c$ can be conveniently written as
\begin{equation}\label{eq:kappa_eta}
  \kappa_c =  \frac{\pi^2  T n_c}{3 m^*_c} \tau^\kappa_c,\quad
  \eta_c =  \frac{ p_{{\rm F}c}^2 n_c}{5 m^*_c} \tau^\eta_c,
\end{equation}
where $n_c$ is the number density of the corresponding species,
$p_{\mathrm{F}c}$ is their Fermi momentum, and $m_c^*$ is their
effective mass on the Fermi surface. The quantities
$\tau^{\kappa,\eta}_c$ are  effective relaxation times which are
generally not the same for different transport problems (thermal
conductivity and shear viscosity in present case, as indicated by
the corresponding superscripts here and in the rest of the paper)
and need to be determined from the transport theory.

The effective relaxation times $\tau^{\kappa,\eta}_c$  are found
from the solution of a system of coupled transport equations.
However, for strongly degenerate matter in NS cores it is enough
to rely on the simplest variational solution of this system
\cite{ShterninYakovlev2007PhRvD,ShterninYakovlev2008PhRvD,SchmittShternin2017arXiv}
(see, however, Sec.~\ref{sec:corr_Var}). Then the problem of
finding effective relaxation times reduces to a system of
algebraic equation
\begin{equation}\label{eq:collfreq_system}
  1=\sum_i \left(\nu_{ci} \tau_{c}+\nu_{ci}'\tau_i\right),
\end{equation}
where indices $c$, $i$ number particles species and the effective
collision frequencies $\nu_{ci}$ and $\nu'_{ci}$ are related to
the transport cross-sections as shown below. The correction to the
variational solution for lepton transport coefficients in normal
matter was found to be within 10\%
\cite{ShterninYakovlev2007PhRvD,ShterninYakovlev2008PhRvD} which
is unimportant for practical applications. The frequencies
$\nu_{ci}$ describe relaxation due to collisions of particle
species $c$ with all other particles including the passive
scatterers. The primed quantities $\nu_{ci}'$ are the mixing
terms. Notice, that the summation in
Eq.~(\ref{eq:collfreq_system}) is carried over all particle
species in both terms, so that the actual collision frequency for
collisions of like particles is $\nu_{cc}+\nu_{cc}'$. These two
parts are kept separated for convenience.

Collision frequencies are calculated by integrating  the squared
matrix element $|M_{ci}|^2$ of corresponding interaction over the
available phasespace with certain phase factors. Consider particle
collisions $c,i\to c',i'$. Primes here mark the particle states
after the collision. Due to a strong degeneracy, the particle
states before and after the collision can be placed on the
respective Fermi surfaces whenever possible, hence the absolute
values of input and output momenta are fixed:
$p_c=p_{c'}=p_{\mathrm{F}c}$ and $p_i=p_{i'}=p_{\mathrm{F}i}$.
Owing to the momentum conservation, the relative orientation of
the four participating momenta is fixed by two angular variables.
In case of electromagnetic collisions, the convenient pair of
variables is the absolute value of the transferred momentum $q$,
where $\mathbf{q}=\mathbf{p}_{c'}-\mathbf{p}_c$, and the angle
$\phi$ between the vectors $\mathbf{p}_c+\mathbf{p}_{c'}$ and
$\mathbf{p}_i+\mathbf{p}_{i'}$. Notice, that these two vectors are
transverse to $\mathbf{q}$.  It is instructive to introduce the
spin-averaged squared matrix element ${\cal
Q}_{ci}(\omega,q,\phi)=(1+\delta_{ci})^{-1}\sum_\mathrm{spins}|M_{ci}|^2/4$,
where the factor $(1+\delta_{ci})^{-1}$ is included in order to
avoid double counting of the same collisions when antisymmetrized
amplitudes are used. In general, ${\cal Q}_{ci}$ depends also on
the transferred energy $\omega=\epsilon_{c'}-\epsilon_c$, where
$\epsilon_{c}$ is the particle energy. In degenerate matter,
$\omega$ is of the order of $T$ and therefore small. In the limit
$\omega\ll q v_{\mathrm{F}i}$, the collision frequencies to be
used in (\ref{eq:collfreq_system}) are
\cite{ShterninYakovlev2007PhRvD,ShterninYakovlev2008PhRvD}
\begin{eqnarray}
\nu^\kappa_{ci}&=&\frac{3 T^2 m_c^*m_i^{*2}}{4\pi^4 p_{\mathrm{F}c} }\nonumber\\
&&\times
\left\langle\frac{\omega^2}{\pi^2 T^2}
\left(1+\left[\frac{\pi^2 T^2}{3\omega^2}-\frac{1}{6}\right]\frac{q^2}{p_{\mathrm{F}c}^2}\right)\,{\cal Q}_{ci}\right\rangle,\label{eq:nu_kappa}
\end{eqnarray}
\begin{eqnarray}
{\nu'}^{\kappa}_{ci}&=&-\frac{3 T^2 p_{\mathrm{F}i} m_c^{*2}m_i^*}{4\pi^4 p_{\mathrm{F}c}^2}\label{eq:nu_kappa1} \\
&&\times
\left\langle\frac{\omega^2}{\pi^2 T^2}
\sqrt{
\left(1-\frac{q^2}{4p_{\mathrm{F}c}^2}\right)
\left(1-\frac{q^2}{4p_{\mathrm{F}i}^2}\right)}{ \cos\phi}\, {\cal Q}_{ci}\right\rangle,\nonumber
\end{eqnarray}
\begin{equation}
\nu^\eta_{ci}=\frac{3 T^2 m_c^* m_i^{*2}}{4\pi^4 p_{\mathrm{F}c} }
\left\langle{\frac{q^2}{p_{\mathrm{F}c}^2}}
\left(1-\frac{q^2}{4p_{\mathrm{F}c}^2}\right)\,{\cal Q}_{ci}\right\rangle,\label{eq:nu_eta}
\end{equation}
\begin{eqnarray}
{\nu'}^{\eta}_{ci}&=&-\frac{3 T^2 p_{\mathrm{F}i} m_c^{*2}m_i^*}{4\pi^4 p_{\mathrm{F}c}^2}\label{eq:nu_eta1}\\
&&\times
\left\langle{\frac{q^2}{p_{\mathrm{F}c}^2}}
\sqrt{\left(1-\frac{q^2}{4p_{\mathrm{F}c}^2}\right)
\left(1-\frac{q^2}{4p_{\mathrm{F}i}^2}\right)} {\cos\phi}\, {\cal Q}_{ci}\right\rangle,\nonumber
\end{eqnarray}
where the angular brackets denote phase-space integration
\begin{equation}\label{eq:phase_aver}
\langle\cdot\rangle=\int\limits_0^\infty d w\,
\frac{\left(w/2\right)^2}{\mathrm{sinh^2
}\left(w/2\right)}\int\limits_0^{q_m} d q \int\limits_0^\pi
d \phi\,\cdot\, ,
\end{equation}
$w=\omega/T$, and
$q_m=\mathrm{min}(2p_{\mathrm{F}c},2p_{\mathrm{F}i})$. Dependence
of ${\cal Q}_{ci}$ on $\omega$ determines the temperature behavior
of collision frequencies and hence of the corresponding transport
coefficients. In traditional transport theory of Fermi systems,
the transition probability is assumed to be independent of
$\omega$. Then each collision frequency in
Eqs.~(\ref{eq:nu_kappa})--(\ref{eq:nu_eta1}) obeys
$\nu_{ci}\propto T^2$ scaling which according to
Eqs.~(\ref{eq:kappa_eta})--(\ref{eq:collfreq_system})  results in
standard dependencies $\kappa\propto T^{-1}$ and $\eta\propto
T^{-2}$. These relations hold, for instance, for the transport
coefficients in the nucleon sector, e.g.
\cite{SchmittShternin2017arXiv}.

Consider leptonic (electrons and muons) subsystem. Leptons collide
with all charged particles due to electromagnetic interaction.
The matrix element of this interaction can be written as a sum of
the longitudinal and transverse parts
\begin{equation}\label{eq:CoulombMatel}
M_{ci}=4\pi \alpha_f \left(\frac{J^{(0)}_c J^{(0)}_i }{q^2+\Pi_l(\omega,q)}-\frac{\mathbf{J}_{c,t}\cdot\mathbf{J}_{i,t}}{q^2-\omega^2+\Pi_t(\omega,q)}\right)\, ,
\end{equation}
where $\alpha_f\approx 1/137$ is the fine structure constant,
$J^{(0)}_c$ and $\mathbf{J}_{c,t}$ are time-like and transverse
(with respect to $\mathbf{q}$) space-like components of the
transition current, respectively, and $\Pi_l$ and $\Pi_t$ are the
longitudinal and transverse polarization functions, respectively.

The transition four-current in Eq.~(\ref{eq:CoulombMatel}) is
$J^\lambda_c=Z_c \bar{u}(\mathbf{p}_{c'})\gamma^\lambda
u(\mathbf{p}_c)/(2\sqrt{\epsilon_c\epsilon_{c'}})$, where $Z_c$ is
the charge number of the particle species $c$, $\gamma^\lambda$ is
a Dirac matrix, and $u(\mathbf{p}_c)$ is the Dirac spinor.
Performing spin summations (in the limit $\omega\ll q
v_{\mathrm{F}i}$), one obtains
\cite{ShterninYakovlev2007PhRvD,ShterninYakovlev2008PhRvD,Alford:2014doa}
\begin{widetext}
\begin{equation}\label{eq:Qci}
{\cal Q}_{ci}=16\pi^2\alpha_f^2Z_c^2 Z_i^2
\left(\frac{L_l}{\left|q^2+\Pi_l(\omega,q)\right|^2}
-2\mathrm{Re} \frac{v_{\mathrm{F}c} v_{\mathrm{F}i} L_{tl}}{(q^2+\Pi_l(\omega,q))(q^2+\Pi_t(\omega,q))^*}
+\frac{v_{\mathrm{F}c}^2 v_{\mathrm{F}i}^2 L_t}{\left|q^2+\Pi_t(\omega,q)\right|^2}\right),
\end{equation}
\end{widetext}
where the numerators are
\begin{eqnarray}
L_l&=&\left(1-\frac{q^2}{4m_c^{*2}}\right)\left(1-\frac{q^2}{4m_i^{*2}}\right),\label{eq:Ll}\\
L_{tl}&=&\sqrt{\left(1-\frac{q^2}{4p_{\mathrm{F}c}^2}\right)\left(1-\frac{q^2}{4p_{\mathrm{F}i}^2}\right)}{\cos\phi},\label{eq:Ltl}\\
L_t&=&\left(1-\frac{q^2}{4p_{\mathrm{F}c}^2}\right)\left(1-\frac{q^2}{4p_{\mathrm{F}i}^2}\right){\cos^2\phi}\nonumber\\
&&+\frac{q^2}{4p_{\mathrm{F}c}^2}+\frac{q^2}{4p_{\mathrm{F}i}^2}.\label{eq:Lt}
\end{eqnarray}
In case of identical particles, ${\cal Q}_{cc}$  also contains an
exchange contribution from the interference between two scattering
channels with the final states interchanged. However, for the
electromagnetic collisions, small momentum transfer $q\ll
p_{\mathrm{F}c}$ dominates the scattering,  interference
corrections  are of the next order in $q$ and are found to be
negligible
\cite{ShterninYakovlev2007PhRvD,ShterninYakovlev2008PhRvD}.

As follows from Eq.~(\ref{eq:Qci}), ${\cal Q}_{ci}$ has
contributions from longitudinal, transverse, and mixed parts of
electromagnetic interaction. Moreover, due to a specific
$\cos\phi$ dependence in
Eqs.~(\ref{eq:nu_kappa})--(\ref{eq:nu_eta1}) and
(\ref{eq:Ll})--(\ref{eq:Ltl}), the mixed term does not contribute
to `direct' collision frequencies (\ref{eq:nu_kappa}) and
(\ref{eq:nu_eta}), so one can write
$\nu_{ci}=\nu_{ci}^l+\nu_{ci}^t$. In contrast, only the mixed term
contributes to the primed collision frequencies,
$\nu_{ci}'=\nu_{ci}^{tl}$
\cite{ShterninYakovlev2007PhRvD,ShterninYakovlev2008PhRvD}. In the
non-relativistic limit $v_{\mathrm{F}c/i}\ll 1$ and the transverse
part of the interaction is unimportant. However, it turns out that
for relativistic particles this part gives the dominant
contribution because of the weaker screening. The leading $q^{-4}$
dependence of ${\cal Q}_{ci}$ is regularized at small $q$ by the
polarization functions $\Pi_l$ and $\Pi_t$ which play the central
role in determining the collision frequencies. Characters of the
longitudinal and transverse screening are very different.  It is
enough to consider screening in the limits of small $q\ll
p_{\mathrm{F}i}$ and $\omega\ll \mu_i$, where $\mu_i$ is the
chemical potential of the i species, and also in the static limit
$ \omega \ll q v_{\mathrm{F}i}$. Then the longitudinal part of the
interaction is screened on a static Thomas-Fermi scale
\begin{equation}\label{eq:Pil_norm}
\Pi_l(\omega,q)= q_\mathrm{TF}^2\equiv \frac{4\alpha_f}{\pi}\sum_i Z_i^2 m_i^* p_{\mathrm{F}i},
\end{equation}
where $q_\mathrm{TF}$ is the Thomas-Fermi screening momentum. In
contrast, the transverse screening is dynamical
\begin{equation}\label{eq:Pit_norm}
\Pi_t(\omega,q)=i\;\frac{\pi}{4}\frac{\omega}{q}q_{t}^2\equiv i\; \frac{\omega}{q} \alpha_f\sum_i Z_i^2 p^2_{\mathrm{F}i},
\end{equation}
where $q_t$ is a characteristic transverse momentum. Therefore the
screening scale of the transverse part of the interaction is $\sim
(\omega q_t^2)^{1/3} \ll q_\mathrm{TF}$ [examine the denominator
in the third term in Eq.~(\ref{eq:Qci})]. This leads to dominant
contribution of the transverse interaction to the collision
frequencies, $\nu_{ci}^t\gg \nu_{ci}^l,\,\nu_{ci}^{tl}$. As a
consequence, the system (\ref{eq:collfreq_system}) decouples, and
in the leading order $\tau_c=\left(\sum_i \nu_{ci}^t\right)^{-1}$.
Retaining only the transverse contribution and the leading order
in $q$ in Eqs.~(\ref{eq:nu_kappa})--(\ref{eq:nu_eta1}), one gets
the following expressions for the lepton thermal conductivity and
shear viscosity in normal matter
\cite{ShterninYakovlev2007PhRvD,ShterninYakovlev2008PhRvD,SchmittShternin2017arXiv}
\begin{eqnarray}
\kappa_{e\mu}&=&\frac{\pi^2}{54\zeta(3)}\frac{p_{\mathrm{F}e}^2+p_{\rm F\mu}^2}{\alpha_f},\label{eq:kappa_emu_trans_norm} \\
\eta_{e\mu} &=&\frac{1.1}{\alpha_f}
\frac{n_e^2+n_\mu^2}{q_t^{1/3}} {T^{-5/3}},\label{eq:eta_emu_trans_norm}
\end{eqnarray}
where $\zeta(3)$ is the Riemann zeta-function. Notice the unusual
temperature behavior of $\kappa_{e\mu}$ and $\eta_{e\mu}$ in
comparison to the standard Fermi-liquid results. This is a
consequence of the dynamical character of the transverse
screening. The different powers of $T$ in
Eqs.~(\ref{eq:kappa_emu_trans_norm}) and
(\ref{eq:eta_emu_trans_norm}) are traced back to the different
leading orders in $q$ for the thermal conductivity ($q^0$) and
shear viscosity ($q^2$) problems in
Eqs.~(\ref{eq:nu_kappa})--(\ref{eq:nu_eta1}). Expression
(\ref{eq:kappa_emu_trans_norm}) is a good approximation to the
exact result, which includes all contributions to collision
frequencies. For the shear viscosity, the dominance of the
transverse part of interaction  is not so strong, and
Eq.~(\ref{eq:eta_emu_trans_norm}) can actually result in a strong
overestimation of the shear viscosity coefficient
\cite{ShterninYakovlev2008PhRvD,Kolomeitsev2015PhRvC}. In this
case all terms need to be retained.

\section{Lepton transport coefficients in superconducting NS cores}\label{sec:transport_sc}
\citet{ShterninYakovlev2007PhRvD,ShterninYakovlev2008PhRvD} also
calculated $\kappa_{e\mu}$ and $\eta_{e\mu}$ in the case when the
protons are in the paired state. They noticed that the proton
pairing changes the character of transverse plasma screening from
the dynamical to the static one restoring the Fermi-liquid
behavior of transport coefficients. Below I show that this
qualitative result is correct, but the treatment of screening in
Refs.~\cite{ShterninYakovlev2007PhRvD,ShterninYakovlev2008PhRvD}
was incomplete. For simplicity, I restrict myself to the case of
well-developed superconductivity $T\lesssim 0.2 \Delta$. For
$^1S_0$ pairing, dependence of the superfluid gap on temperature
can be approximated as \cite{Levenfish1994ARep}
\begin{equation}\label{eq:gap}
\frac{\Delta}{T}=\sqrt{1-t}\left(1.456-\frac{0.157}{\sqrt{t}}+\frac{1.764}{t}\right),
\end{equation}
where $t=T/T_{\mathrm{c}p}$. Thus the condition $T\lesssim 0.2
\Delta$ translates to  $T/T_{\mathrm{c}p}\lesssim 0.35$.  In this
case, first of all,  lepton-proton collisions can be neglected,
and, second, the zero-temperature limit for the proton
polarization function can be used. Provided high expected values
of $T_{\mathrm{c}p}$, this limit is comfortably satisfied at
$T\lesssim 10^8$~K.
\subsection{Plasma screening in presence of proton pairing}\label{sec:prot_screen}

Pairing of protons, which are charged particles, modifies the
screening of the electromagnetic interaction. Up to the order
$\Delta/\mu_p$, the static longitudinal screening does not change
\cite{Gusakov2010PhRvC, Arseev2006PhyU} and $\Pi_l$ is given by
Eq.~(\ref{eq:Pil_norm}). In contrast, the transverse screening
modifies. At low $T$, the dominant contribution to screening comes
from protons. In the  zero-temperature limit $T\to 0$ in the
Bardeen-Cooper-Schrieffer (BCS) theory the static ($\omega\to 0$)
transverse polarization function  can be written as
\cite{Landau9eng}
\begin{equation}\label{eq:Pit_static}
\Pi_t(0,q)=q_M^2 J(\zeta),
\end{equation}
where $\zeta= q v_{\mathrm{F}p}/\Delta=\pi q\xi$, $\xi$ is the
coherence length, and $q_M$ is the Meissner screening momentum
(Meissner mass)
\begin{equation}\label{eq:Meissner}
 q_M^2=\frac{4\alpha
_f}{3\pi} p_{\mathrm{F}p}^2 v_{\mathrm{F}p}.
\end{equation}
The function $J(\zeta)$ in Eq.~(\ref{eq:Pit_static}) is \cite{Landau9eng}
\begin{eqnarray}\label{eq:J_T0}
J(\zeta)&=&\frac{3}{8}\int\limits_{-\infty}^{\infty} d \eta \int\limits_{-1}^1 \frac{d x\, (1-x^2)}{\sqrt{\eta^2+1} \left(\eta^2+1+(\zeta x)^2/4\right)}\nonumber\\
&=&\frac{3}{4}\int\limits_{-1}^{1} d x\, (1-x^2)\, \frac{2{\rm ArcSinh}(\zeta x/2)}{\zeta x \sqrt{1+(\zeta x)^2/4}}.
\end{eqnarray}
In principle, the integration over  $x$  in the second line in
Eq.~(\ref{eq:J_T0}) can be performed analytically with the result
being expressed via the polylogarithmic functions.

\begin{figure}[ht]
\includegraphics[width=0.9\columnwidth]{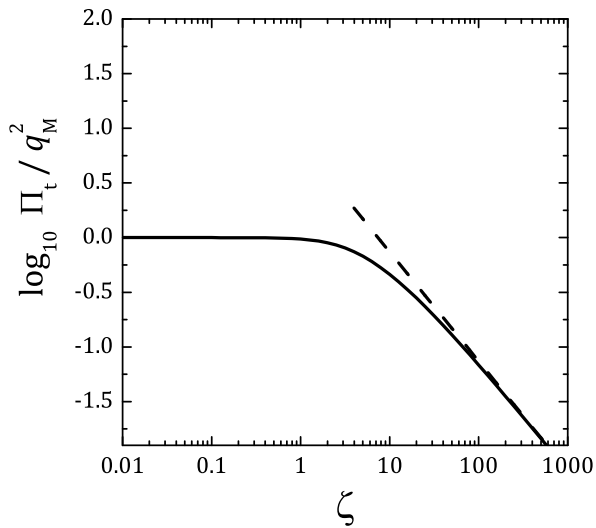}
\caption{Proton contribution to the zero-temperature transverse
polarization function in the static limit. Polarization function
is normalized to $q_M^2$. Dashed line shows the asymptote in the
Pippard limit.}\label{fig:Pi}
\end{figure}
The function $J(\zeta)$ is plotted in Fig.~\ref{fig:Pi}. At small
$\zeta$ (small momentum $q\ll \xi^{-1}$), which corresponds to the
London limit, $J(\zeta)=1$. In this limit, the transverse
screening is independent of $\Delta$ and the screening momentum is
equal  to $q_M$. The real transverse photons obey the Meissner
mass $q_M$ in this limit. This leads to the Meissner effect in
superconductors. In the opposite, Pippard limit, $\zeta\gg 1$ and
$\Pi_t$ is inversely proportional to $\zeta$ as shown by the
dashed line in  Fig.~\ref{fig:Pi}. The asymptotic expression in
the Pippard limit reads  $J(\zeta)=3\pi^2/(4\zeta)$. In this
limit, the characteristic transverse screening momentum is
$q_P=(3\pi^2 \Delta q_M^2/(4v_{\mathrm{F}p}))^{1/3}$. This
expression resembles the screening momentum in the non-superfluid
case, with $\Delta$ in place of $\omega$ and $q_M$ in place of
$q_t$. In the Pippard limit, contrary to the London limit, the
screening depends on $\Delta$.

\begin{figure}[ht]
\includegraphics[width=0.9\columnwidth]{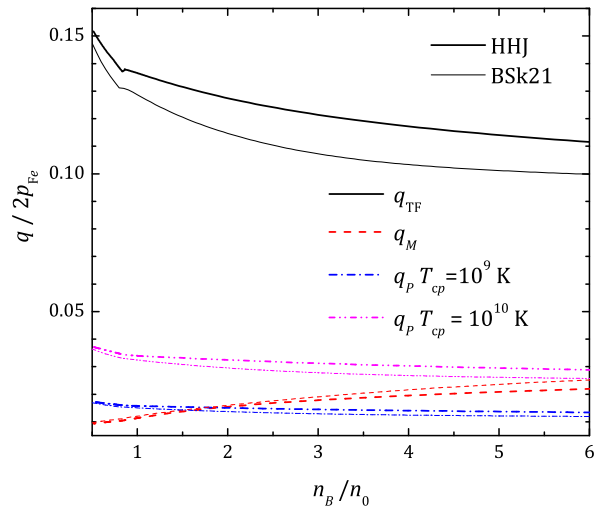}
\caption{(Color online). Ratios of characteristic screening momenta to $2p_{\mathrm{F}e}$ versus baryon density (in units of the nuclear saturation density $n_0=0.16~{\rm fm}^{-3}$) for two EOSs desribed in the text. Thick lines correspond to the HHJ EOS, while thin lines show the results for the BSk21 EOS. Solid lines show the longitudinal (Thomas-Fermi) screening momentum $q_\mathrm{TF}$, dashed lines give the Meissner momentum $q_M$. Dash-dotted and double-dot-dashed lines show the characteristic screening momenta in Pippard limit for $T_{\mathrm{c}p}=10^9$~K and $10^{10}$~K, respectively.}\label{fig:qAPR}
\end{figure}

In Fig.~\ref{fig:qAPR}, the characteristic transverse screening
momenta $q_M$ (dashed lines) and $q_P$ (dash-dotted lines for
$T_{\mathrm{c}p}=10^9$~K and double-dot-dashed lines for
$T_{\mathrm{c}p}=10^{10}$~K) are compared with the longitudinal
screening momentum $q_\mathrm{TF}$ (solid lines). The momenta in
the plot are normalized to  $2p_{\mathrm{F}e}$, which is the
maximum momentum transfer in electron-electron collisions. Thick
and thin lines correspond to two widely used equations of state
(EOSs) of  dense nucleon matter in NS cores. Namely, by the
abbreviation HHJ (thick lines) I denote the EOS constructed by
\citet{Heiselberg1999ApJ} as an analytical parameterization of the
variational EOS by \citet{Akmal1998PhRvC}. Specifically, I use the
model with the parameter $\gamma=0.6$ of
Ref.~\cite{Heiselberg1999ApJ}; this model was designated as APR~I
in Ref.~\cite{Gusakov2005MNRAS} and the NS properties with such
EOS can be found there. With thin lines I show the results for one
of the EOSs based on the Brussels-Skyrme nucleon interaction
functionals, namely the BSk21 model \cite{Potekhin2013A&A}. Both
EOSs satisfy the equilibrium conditions with respect to the weak
processes. Unless otherwise indicated, the proton effec1tive mass
is set to $m_p^*=0.8 m_u$, where $m_u$ is the nucleon mass unit.
Two EOSs are different in the particle fractions, however the
results shown in Fig.~\ref{fig:qAPR} are qualitatively same. As in
the normal matter, characteristic transverse screening momenta
($q_M$ or $q_P$) are much smaller than the longitudinal one. As a
consequence, the transverse part of the interaction dominates in
the presence of proton superconductivity as well.

\begin{figure}[ht]
\includegraphics[width=0.8\columnwidth]{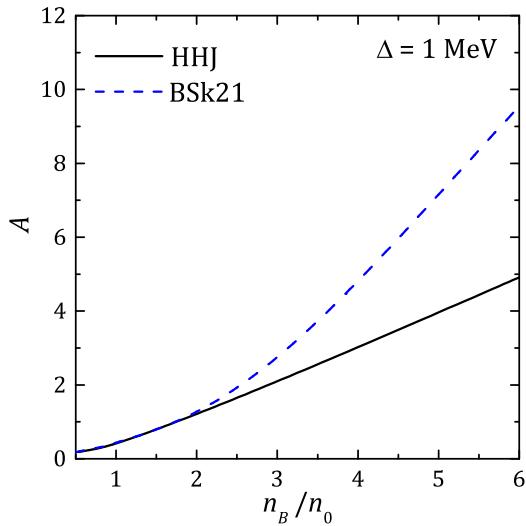}
\caption{(Color online). The coherence scale parameter $A$ as a
function of the total baryon density  for two selected EOSs.
Effective masses are set to $m^*_p=0.8 m_u$ and
$\Delta=1$~MeV.}\label{fig:A}
\end{figure}

In
Refs.~\cite{ShterninYakovlev2007PhRvD,ShterninYakovlev2008PhRvD}
it was assumed that the typical transferred momentum $q$ is not so
small, so that the Pippard limit gives appropriate description of
the transverse plasma screening in NS cores. In fact, which limit,
London or Pippard, gives the dominant contribution depends on the
value of $\zeta$ at $q=q_M$. This point was overlooked in
Refs.~\cite{ShterninYakovlev2007PhRvD,ShterninYakovlev2008PhRvD}.
It is hence convenient to introduce the parameter
$A\equiv\zeta(q=q_M)=v_{\mathrm{F}p} q_M/\Delta$. In the BCS
approximation, this parameter is related to the familiar
Ginzburg-Landau coherence parameter $\varkappa$, namely
$\varkappa=\lambda_L/\xi=\pi/A$, where $\lambda_L=q_M^{-1}$ is the
London penetration depth.  The value of $\varkappa$ determines the
superconductivity type. The transition from type I superconductor
to type II superconductor occurs at  $\varkappa>1/\sqrt{2}$ as
type-I  \cite{TilleyTilley1990,Landau9eng}, which corresponds to
$A< \sqrt{2}\pi\approx 4.4$.\footnote{Notice, that this criterion
modifies in the superfluid-superconducting mixtures
\cite{HaberSchmitt2017PhRvD} which is quite possible in NS cores
where (in large part at least) neutrons can also be in the
superfluid state. According to Ref.~\cite{HaberSchmitt2017PhRvD},
the point $\varkappa=1/\sqrt{2}$ does not separate the
topologically different type-I and type-II phases in this case,
and situation is more complicated. Since the results of the
present paper are not affected by these complications, we will
nevertheless call the region where $\varkappa>1/\sqrt{2}$ as
type-II superconductivity region, and where $\varkappa<1/\sqrt{2}$
as type-I superconductivity region for simplicity.} The parameter
$A$ can be written as
\begin{equation}\label{eq:A_natural}
A=1.12 \left(\frac{x_p}{0.1}\right)^{5/6}\left(\frac{n_B}{n_0}\right)^{5/6} \left(\frac{m^*_p}{m_u}\right)^{-3/2} \frac{0.5~{\rm MeV}}{\Delta},
\end{equation}
where $n_0=0.16~{\rm fm}^{-3}$ is the nuclear saturation density.
In Fig.~\ref{fig:A}, the parameter $A$ is plotted for two  EOSs
discussed above and for $\Delta = 1$~MeV. This corresponds to
$T_{\mathrm{c}p}\approx 6.5\times 10^9$~K [see Eq.~(\ref{eq:gap})
at $T=0$]. For this large $\Delta$, most of the core forms type-II
superconductor~\cite{Baym1969Nature}. Since $A$ is inversely
proportional to $\Delta$, it is higher for lower $\Delta$ (lower
$T_{\mathrm{c}p}$). For the NS core conditions, $A$ can vary in
the range $0.1\div 100$. Figure~\ref{fig:Pi} shows that these
values correspond to the intermediate region between the London
and Pippard limits, thus one can expect that neither of these
limits is strictly applicable in NS cores, and the general form of
$\Pi_t$ should be used for calculating the transport coefficients.
This is demonstrated in the next Section.

\subsection{Calculation of transport coefficients in the leading order}\label{sec:collfreq_lead}

According to the discussion in Secs.~\ref{sec:kin_form} and
\ref{sec:prot_screen} (see also Fig.~\ref{fig:qAPR}), the dominant
contribution to the lepton collision frequencies comes from the
transverse part of the electromagnetic interaction. In addition,
since the screening is weak, the lowest order in $q$ in
Eqs.~(\ref{eq:nu_kappa})--(\ref{eq:nu_eta1}),
(\ref{eq:Qci})--(\ref{eq:Lt})  gives the leading contribution to
transport coefficients. In order to calculate the transverse
collision frequencies $\nu_{ci}^t$, the integrals
\begin{equation}\label{eq:In}
I^t_n(A,q_M)=\int_0^{q_m}\frac{dq \, q^n}{(q^2+q_M^2 J(\zeta))^2}
\end{equation}
are needed. The exponent $n=0$ in Eq.~(\ref{eq:In}) gives the
leading order contribution for the thermal conductivity problem,
while for the shear viscosity the leading order is given by $n=2$
(Sec.~\ref{sec:kin_form}). Retaining only the leading
contributions one gets the following results for the lepton
thermal conductivity and shear viscosity
\begin{eqnarray}
\kappa_{e\mu}^{t,\,\mathrm{Lead}} &=& \frac{5}{72\pi\alpha^2_f { T}} \left[I^t_0(A,q_M)\right]^{-1},\label{eq:kappa_lead}\\
\eta_{e\mu}^{t,\,\mathrm{Lead}} &=&\frac{3\pi  }{10\alpha_f^{2} T^2}
\frac{n_e^2+n_\mu^2}{p_{\mathrm{F}e}^2+p_{\mathrm{F}\mu}^2}
\left[I^t_2(A,q_M)\right]^{-1}.\label{eq:eta_lead}
\end{eqnarray}

Let us analyze an asymptotic behavior of the integrals
(\ref{eq:In}). In the weak-screening limit $q_M\ll q_m$ it is
enough to extend the upper integration limit to infinity. Then,
the low-$A$ asymptote becomes
\begin{equation}\label{eq:In_small}
I^t_n=\frac{\pi}{4 q_M^{3-n}},\quad A\ll 1,
\end{equation}
while the high-$A$ asymptotes are
\begin{equation}\label{eq:In_large}
I^t_0=\frac{4A}{9\pi^2q_M^3},\quad I^t_2=\frac{4A^{1/3}}{9 q_M}\, \frac{2^{2/3} \pi^{1/3}}{3^{5/6}},\quad A\gg 1.
\end{equation}
Remarkably, the low-$A$ asymptote (\ref{eq:In_small}), which
corresponds to the London limit, is independent of $A$ and hence
of $\Delta$. This is a consequence of the independence of the
Meissner momentum $q_M$ of the gap value. The case of large $A$,
Eq.~(\ref{eq:In_large}), corresponds to the Pippard limit that was
employed in
Refs.~\cite{ShterninYakovlev2007PhRvD,ShterninYakovlev2008PhRvD}.
In the intermediate case, the integrals $I_0^t$ and $I_2^t$ were
fitted by the analytic expressions to facilitate their use in
applications. These expressions are given in the Appendix.
Substituting the limiting expressions
(\ref{eq:In_small})--(\ref{eq:In_large}) into
Eqs.~(\ref{eq:kappa_lead})--(\ref{eq:eta_lead}), one obtains the
asymptotic expressions for the thermal conductivity and shear
viscosity
\begin{eqnarray}
\kappa_{e\mu}^\mathrm{Lon} &=& \frac{5 q_M^3}{18\pi^2\alpha_f^2 T},\label{eq:kappa_emu_Meiss}\\
\eta_{e\mu}^\mathrm{Lon} &=&\frac{6 q_M}{5\alpha_f^{2}\, T^2}
\frac{n_e^2+n_\mu^2}{p_{\mathrm{F}e}^2+p_{\mathrm{F}\mu}^2},
\label{eq:eta_emu_Meiss}\\
\kappa_{e\mu}^\mathrm{Pip} &=& \frac{5\,p_{\mathrm{F}p}^2}{24\alpha_f}
\frac{ \Delta}{{ T}},\label{eq:kappa_emu_Pipp}\\
\eta_{e\mu}^\mathrm{Pip} &=&\frac{1.71\,p_{\mathrm{F}p}}{\alpha_f^{5/3}\,T^2}
\frac{n_e^2+n_\mu^2}{p_{\mathrm{F}e}^2+p_{\mathrm{F}\mu}^2}
{ \left(\frac{\Delta}{p_{\mathrm{F}p} c}\right)^{1/3}},\label{eq:eta_emu_Pipp}
\end{eqnarray}
where the superscripts Lon and Pip correspond to the London and Pippard approximations to $\Pi_t$, respectively.
\begin{figure}[ht]
\includegraphics[width=0.9\columnwidth]{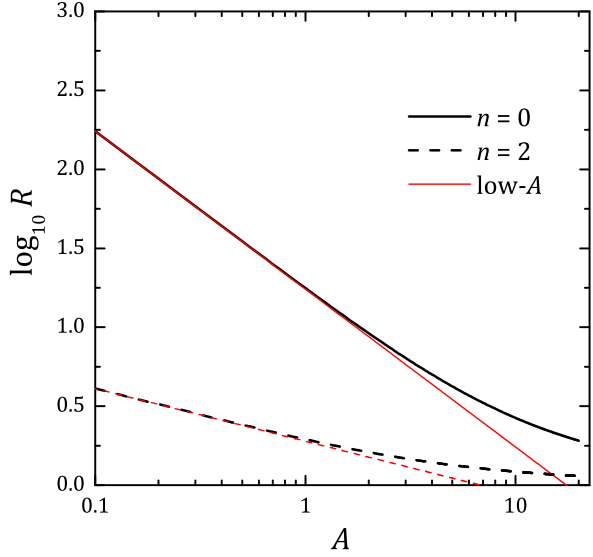}
\caption{(Color online). The ratio $R(A)$  for $n=0$ (solid lines)
and $n=2$ (dashed lines). Thin lines represent the low-$A$
approximation (\ref{eq:In_small}).}\label{fig:R-A}
\end{figure}

Comparing the expressions (\ref{eq:In_small}) and
(\ref{eq:In_large}) one can roughly estimate that the crossover
between the two limiting cases occurs at $A\approx 17.5$ for $n=0$
and at $A\approx 6.8$ for $n=2$. From Fig.~\ref{fig:A}, one
concludes that for large $\Delta\sim 1$~MeV or for small $n_B$ if
$\Delta$ is lower, the Pippard limit used in
Refs.~\cite{ShterninYakovlev2007PhRvD,ShterninYakovlev2008PhRvD}
is inapplicable. To illustrate the possible degree of inaccuracy
of the older results, let us construct the ratio $R(A)$  of the
leading contribution to transverse collision frequency to those
calculated in the Pippard limit:
$\nu_t^\mathrm{Lead}=\nu_t^\mathrm{Pip} R(A)$. This ratio is
plotted as a function of $A$ in Fig.~\ref{fig:R-A} for $n=0$
(thermal conductivity, solid lines) and $n=2$ (shear viscosity,
dashed lines).  The plot clearly shows underestimation of the
collision frequencies, and, hence overestimation of the transport
coefficients by the Pippard limiting values
(\ref{eq:kappa_emu_Pipp})--(\ref{eq:eta_emu_Pipp}). For small $A$
and $n=0$ this overestimation reaches two orders of magnitude. For
the shear viscosity problem ($n=2$), the overestimation is modest
because of the weaker dependence of the collision frequencies on
the screening momentum (Sec.~\ref{sec:kin_form}). Thin lines in
Figure~\ref{fig:R-A} show the same factors, where the London
asymptotic expression for collision frequencies is used in place
of $\nu^{t,\mathrm{Lead}}$. One concludes, that the London limit
for screening is appropriate in the case of type-II
superconductivity, but for large values of $A$ it becomes
inapplicable. Fig.~\ref{fig:R-A} shows that both asymptotic limits
generally underestimate the collision frequencies and overestimate
the transport coefficients. This is because of an overestimation
of the screening at large $q$ by the London expression and at
small $q$ by the Pippard expression, see Fig.~\ref{fig:Pi}.

\begin{figure*}[ht]
\begin{minipage}{0.45\textwidth}
\includegraphics[width=\textwidth]{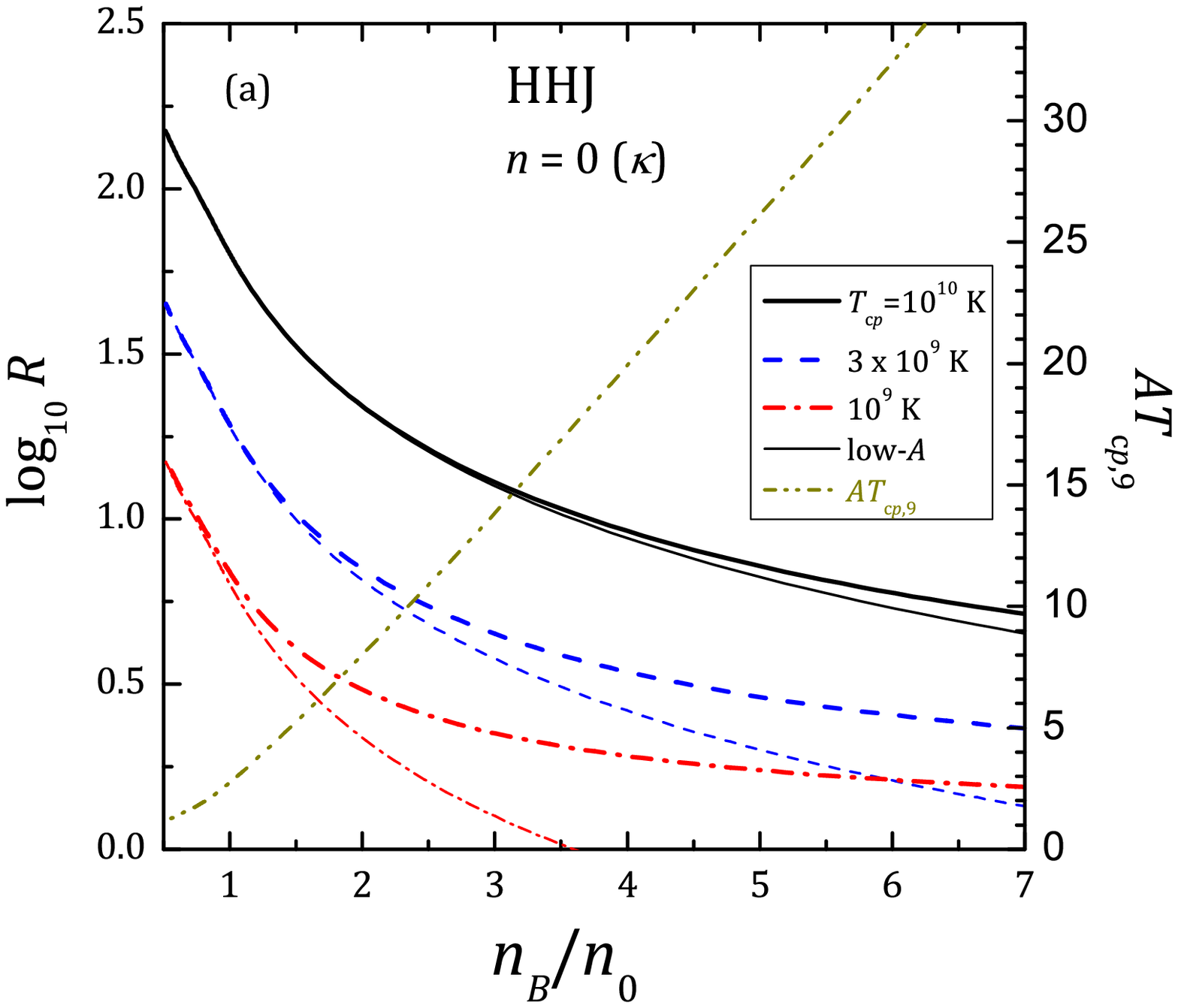}
\end{minipage}
\hspace{0.05\textwidth}
\begin{minipage}{0.45\textwidth}
\includegraphics[width=\textwidth]{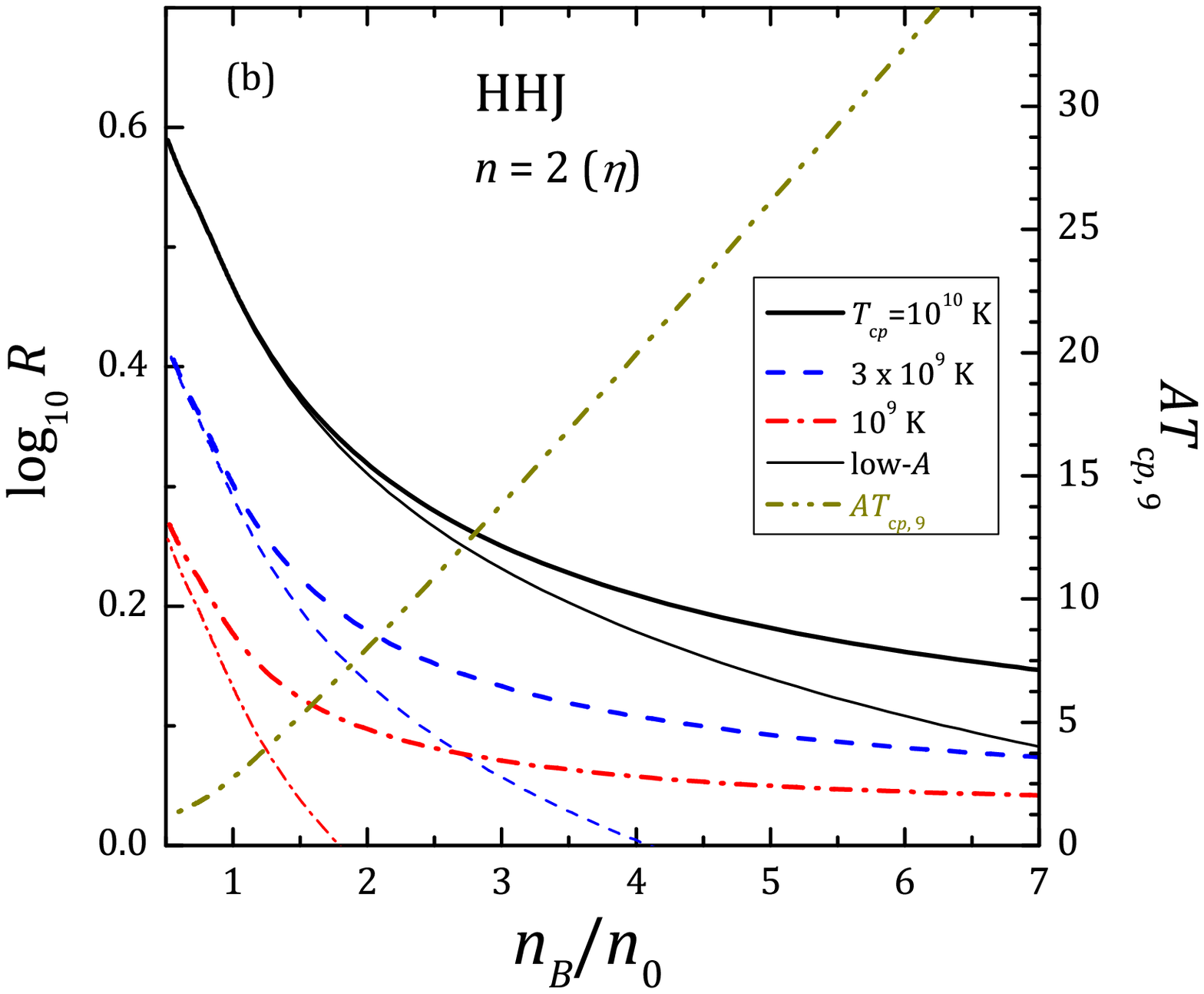}
\end{minipage}
\caption{(Color online). The ratios $R$ for (a) thermal
conductivity ($n=0$) and (b) shear viscosity ($n=2$) as function
of $n_B$ for the HHJ equation of state and three values of
$T_{\mathrm{c}p}=10^{10}$~K (solid lines), $3\times10^9$~K (dashed
lines), and $10^9$~K (dash-dotted lines). Thin lines give low-$A$
approximation, where the interaction is screened by the pure
Meissner mass. Double-dot-dashed lines show with right vertical
scales the combination $AT_{\mathrm{c}p,9}$ which is independent
of $T_{\mathrm{c}p}$. Here $T_{\mathrm{c}p,\,9}\equiv
T_{\mathrm{c}p}/(10^9~\mathrm{K})$.}\label{fig:RAPR}
\end{figure*}

For illustration, the same ratios $R$ are plotted in
Fig.~\ref{fig:RAPR}  now as functions of the baryon density for
the HHJ EOS and three values of $T_{\mathrm{c}p}=10^{10}$~K (solid
lines), $3\times 10^9$~K (dashed lines), and $10^9$~K (dash-dotted
lines). Figure~\ref{fig:RAPR}(a) shows the results appropriate for
the thermal conductivity ($n=0$), while for the shear viscosity
problem ($n=2$), $R$  is shown in Fig.~\ref{fig:RAPR}(b). Like in
Fig.~\ref{fig:R-A}, thin lines give the ratio $R$ calculated with
London limiting expression. For the highest shown critical
temperature, $T_{\mathrm{c}p}=10^{10}$~K,  $R$ is largest and the
London expression is a good approximation for $n=0$ in the whole
shown range of densities and for $n_B\lesssim 3n_0$ for $n=2$.
With decreasing $T_{\mathrm{c}p}$ (increasing $A$), the ratio $R$
lowers down and the applicability range of the London limiting
expression shifts to lower densities. For instance, for
$T_{\mathrm{c}p}=10^{9}$~K and $n=2$ the London approximation is
always inaccurate, as seen from a comparison of thin and thick
dash-dotted lines in  Fig.~\ref{fig:RAPR}(b).

\subsection{Kinematic corrections}\label{sec:collfreq_corr}
\begin{figure}[hb]
\includegraphics[width=0.9\columnwidth]{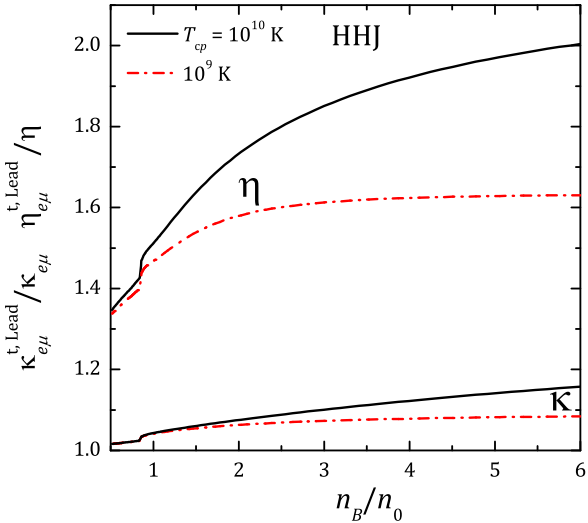}
\caption{(Color online). Comparison of the thermal conductivity
and shear viscosity in the leading approximation,
Eqs.~(\ref{eq:kappa_lead})--(\ref{eq:eta_lead}), to the results of
complete calculations. Upper pair of curves shows the ratio
$\eta^{t,\,\mathrm{Lead}}_{e\mu}/\eta_{e\mu}$, while lower pair of
curves marked $\kappa$ shows
$\kappa^{t,\,\mathrm{Lead}}_{e\mu}/\kappa_{e\mu}$. Results for the
HHJ EOS are shown and for two values of the proton critical
temperature. $T_{\mathrm{c}p}=10^{10}$~K (solid lines) and
$10^9$~K (dashed lines). Variational solutions are employed. See
text for details.}\label{fig:Lead}
\end{figure}
In the previous section the leading order contribution to the
collision frequencies was discussed. However, as was mentioned at
the end of the Sec.~\ref{sec:kin_form}, this approximation can be
inaccurate, especially for the shear viscosity and in principle
the full result that follows from
Eqs.~(\ref{eq:collfreq_system})--(\ref{eq:nu_eta1}) and
(\ref{eq:Qci})--(\ref{eq:Lt}) shall be used
\cite{ShterninYakovlev2008PhRvD,Kolomeitsev2015PhRvC,SchmittShternin2017arXiv}.
The main corrections come from the inclusion of the longitudinal
part of the interaction, and from the kinematic corrections of
high-$q$ powers in   Eqs.~(\ref{eq:nu_kappa})--(\ref{eq:nu_eta1})
and (\ref{eq:Ll})--(\ref{eq:Lt}). Going beyond  the
long-wavelength and static limit in polarization functions is not
necessary, since the possible difference would be sizable at large
$q$, where the $q^2$ term dominates in the denominators in
Eq.~(\ref{eq:Qci}). Since both longitudinal and transverse
screening are static, the  integration over $w$ in
Eqs.~(\ref{eq:nu_kappa})--(\ref{eq:nu_eta1}) can be performed
analytically, as well as the integration over  $\phi$, leaving one
with the following result for the thermal conductivity collision
frequencies
\begin{equation}
\nu^{\kappa}_{ci}=\nu^{\kappa,t}_{ci}+\nu^{\kappa,l}_{ci},\label{nukappa_ci}
\end{equation}
\begin{eqnarray}
\nu^{\kappa,t}_{ci}&=&\frac{8\pi \alpha_f^2 T^2 p_{\mathrm{F}c}p_{\mathrm{F}i}^2}{5m_c^*}\nonumber\\
&&\times\int\limits_0^{q_m}\frac{dq\,\left(1+\frac{q^2}{4p_{\mathrm{F}c}^2}\right)^2\left(1+\frac{q^2}{4p_{\mathrm{F}i}^2}\right)}{\left(q^2+q_M^2J(\zeta)\right)^2},\label{nukappa_ci_tfull}
\end{eqnarray}
\begin{eqnarray}
\nu^{\kappa,l}_{ci}&=&\frac{16\pi \alpha_f^2 T^2 m_c^* m_{i}^{*2}}{5p_{\mathrm{F}c}}\label{nukappa_ci_lfull}\\
&&\times\int\limits_0^{q_m}\frac{dq\,\left(1+\frac{q^2}{4p_{\mathrm{F}c}^2}\right)\left(1-\frac{q^2}{4m_c^{*2}}\right)\left(1-\frac{q^2}{4m_i^{*2}}\right)}{\left(q^2+q_\mathrm{TF}^2\right)^2},\nonumber
\end{eqnarray}
\begin{eqnarray}
{\nu'}^{\kappa}_{ci}&=&\frac{16\pi \alpha_f^2 T^2 m_i^*p_{\mathrm{F}i}}{5}\nonumber\\
&&\times\int\limits_0^{q_m}\frac{dq\,\left(1-\frac{q^2}{4p_{\mathrm{F}c}^2}\right)\left(1-\frac{q^2}{4p_{\mathrm{F}i}^2}\right)}{\left(q^2+q_\mathrm{TF}^2\right)\left(q^2+q_M^2J(\zeta)\right)}\label{nukappa_ci_tlfull}
\end{eqnarray}
and similarly for the shear viscosity collision frequencies
\begin{equation}
\nu^{\eta}_{ci}=\nu^{\kappa,t}_{ci}+\nu^{\kappa,l}_{ci},\label{nueta_ci}
\end{equation}
\begin{eqnarray}
\nu^{\eta,t}_{ci}&=&\frac{2\pi \alpha_f^2 T^2 p_{\mathrm{F}i}^2 }{m_c^* p_{\mathrm{F}c}}\nonumber\\
&&\times\int\limits_0^{q_m}\frac{dq\,q^2\left(1-\frac{q^4}{16p_{\mathrm{F}c}^4}\right)\left(1+\frac{q^2}{4p_{\mathrm{F}i}^2}\right)}{\left(q^2+q_M^2J(\zeta)\right)^2},\label{nueta_ci_tfull}
\end{eqnarray}
\begin{eqnarray}
\nu^{\eta,l}_{ci}&=&\frac{4\pi \alpha_f^2 T^2 m_c^* m_{i}^{*2}}{p_{\mathrm{F}c}^3}\label{nueta_ci_lfull}\\
&&\times\int\limits_0^{q_m}\frac{dq\,q^2\left(1-\frac{q^2}{4p_{\mathrm{F}i}^2}\right)\left(1-\frac{q^2}{4m_c^{*2}}\right)\left(1-\frac{q^2}{4m_i^{*2}}\right)}{\left(q^2+q_\mathrm{TF}^2\right)^2},\nonumber
\end{eqnarray}
\begin{eqnarray}
{\nu'}^{\eta}_{ci}&=&\frac{4\pi \alpha_f^2 T^2 m_i^*p_{\mathrm{F}i}}{p_{\mathrm{F}c}^2}\nonumber\\
&&\times\int\limits_0^{q_m}\frac{d q\,q^2\left(1-\frac{q^2}{4p_{\mathrm{F}c}^2}\right)\left(1-\frac{q^2}{4p_{\mathrm{F}i}^2}\right)}{\left(q^2+q_\mathrm{TF}^2\right)\left(q^2+q_M^2J(\zeta)\right)}.\label{nueta_ci_tlfull}
\end{eqnarray}
Thus in order to calculate the transverse part of the collision
frequencies, $\nu_{ci}^{\kappa,t}$ and $\nu_{ci}^{\eta,t}$,
including all kinematic corrections one needs integrals $I^t_n$
defined in Eq.~(\ref{eq:In}) up to $n=8$. Similarly, to calculate
longitudinal contributions, $\nu_{ci}^{\kappa,l}$ and
$\nu_{ci}^{\eta,l}$,  analogous longitudinal integrals
\begin{equation}\label{eq:Il}
I^l_n(q_\mathrm{TF})=\int_0^{q_m}\frac{d q \, q^n}{(q^2+q_\mathrm{TF}^2)^2}
\end{equation}
are required. These are standard integrals, and their explicit
expressions up to $n=8$ can be found, for instance, in the
Appendix in Ref.~\cite{ShterninYakovlev2008PhRvD}. Finally, to
calculate the mixing terms, we need integrals
\begin{equation}\label{eq:Itl}
I^{tl}_n(A,q_M,q_\mathrm{TF})=\int_0^{q_m}\frac{ d q \, q^n}{\left(q^2+q_{TF}^2\right)\left(q^2+q_M^2J(\zeta)\right)}
\end{equation}
up to $n=6$.

In Fig.~\ref{fig:Lead}, the results of full calculations which are
based on Eqs.~(\ref{nukappa_ci})--(\ref{nueta_ci_tlfull}) and
Eq.~(\ref{eq:collfreq_system}) are compared with the leading-order
results (\ref{eq:kappa_lead})--(\ref{eq:eta_lead}) for two values
of the critical temperature, $T_{\mathrm{c}p}=10^9$~K and
$10^{10}$~K. Clearly, the kinematic corrections to the thermal
conductivity coefficient can be safely ignored in applications.
However, the leading-order expression (\ref{eq:eta_lead})
overestimates the shear viscosity by 50\% for
$T_{\mathrm{c}p}=10^9$ and up to a factor of two for
$T_{\mathrm{c}p}=10^{10}$~K, since in the latter case the
transverse screening momentum is larger, see Fig.~\ref{fig:qAPR}.
It is thus advisable to go beyond the leading-order expression
when calculating $\eta_{e\mu}$. A detailed analysis of various
corrections shows that it is necessary to include all three
contributions -- transverse, longitudinal, and mixed -- but it is
enough to use the lowest-order terms in $q$, namely retain only
$q^2$ in numerators for each of these terms. In this
approximation, $\eta_{e\mu}$ stays within 10\% of the total
result.   The lowest-order contribution to $\nu_{ci}^{\eta,t}$ is
discussed in the previous section, while the explicit
leading-order expression for  $\nu^{\eta,l}_{ci}$ is given in the
Appendix, Eq.~(\ref{eq:nu_eta_ci_l_lead}). It remains to consider
the leading-order contribution to ${\nu'}_{ci}^{\eta}$. Because of
$q^2$ in the numerator and since $q_M\ll q_\mathrm{TF}$, it is
possible to neglect the transverse screening in
Eq.~(\ref{nueta_ci_tlfull}) \cite{ShterninYakovlev2008PhRvD}. Then
the integration over $q$ is  trivial. Explicit result is given in
Eq.~(\ref{eq:nu_eta_ci_tl_lead}). Notice, that this procedure does
not work for ${\nu'}_{ci}^{\kappa}$,
Eq.~(\ref{nukappa_ci_tlfull}). However, as discussed before,
${\nu'}_{ci}^{\kappa}$ is actually not needed.

\subsection{Corrections to variational solution}\label{sec:corr_Var}
Up to now the simplest variational solution of the system of
transport equations was employed. However, it is possible to
obtain the exact solution. For a single-component Fermi liquid,
the general theory was developed in
Refs.~\cite{BrookerSykes1968PhRvL,SykesBrooker1970AnPhy,JensenSmith1968PhLA}
and was extended to the multicomponent case in
Refs.~\cite{FlowersItoh1979ApJ,Anderson1987}. In these references,
the exact solution was given in analytical way in form of the
rapidly converging series. Equivalently, the system of transport
equations can be solved numerically.

Let me briefly outline the method of the exact solution of
transport equations for the thermal conductivity and shear
viscosity problems. Here I mainly follow the notations in
Ref.~\cite{Shternin2017JPhCS}. Instead of
Eq.~(\ref{eq:kappa_eta}), transport coefficients are rewritten in
the form
\begin{equation}\label{eq:kappa_eta_exact}
  \kappa_c =  C^\kappa_c\,\frac{\pi^2  T n_c}{3 m^*_c} \tau_{c0},\quad
  \eta_c =  C^\eta_c\,\frac{ p_{{\rm Fc}}^2 n_c}{5 m^*_c} \tau_{c0},
\end{equation}
where the characteristic relaxation time
\begin{equation}\label{eq:nu_c0}
\tau_{c0}^{-1}=\nu_{c0}=\frac{3 T^2 m_c^* }{8\pi^4 p_{\mathrm{F}c} }\sum\limits_i m_i^{*2}
\left\langle{\cal Q}_{ci}\right\rangle_{q\phi}
\end{equation}
is introduced. By $\langle \cdot \rangle_{q\phi}$ in
Eq.~(\ref{eq:nu_c0}), the $w$-independent part of
Eq.~(\ref{eq:phase_aver}) is denoted. Characteristic relaxation
times are now the same for the thermal conductivity and for the
shear viscosity. The differences between the specific transport
problems are encapsulated in the coefficients $C^\kappa_c$ and
$C^\eta_c$. In order to find these coefficients, one starts from
the system of transport equations for the non-equilibrium
distribution functions $F_c(\mathbf{p}_c)$ for the particle
species c. These equations are then linearized by introducing a
correction to the local equilibrium distribution function as
\begin{equation}\label{eq:F_anzatz}
  F_c(\mathbf{p}_c)=f(x_c)+ \tau_{c0} \Psi_c(x_c)  D(\mathbf{p}_c)\frac{1}{T}
\frac{\partial f(x_c)}{\partial
  x_c},
\end{equation}
where $f(x)=\left[1+\mathrm{exp}(x)\right]^{-1}$ is the
Fermi-Dirac distribution, $x_c=(\epsilon_c-\mu_c)/T$, $\mu_c$ is
the chemical potential and $D(\mathbf{p})$ is the anisotropic part
of the driving term. For the thermal conductivity, $D(\mathbf{p})
= \mathbf{v}\nabla T$ where $\mathbf{v}$ is the particle velocity,
while for the shear viscosity, $D(\mathbf{p})=\left(v_\alpha
p_\beta-3^{-1} v p \delta_{\alpha\beta}\right)
\left(\partial_\beta V_\alpha+ \partial_\alpha V_\beta\right)/2$,
where $\mathbf{V}$ is the hydrodynamical velocity with
$\mathrm{div}\mathbf{V}=0$. Transport coefficients can be found by
substituting Eq.~(\ref{eq:F_anzatz}) into the equations for the
corresponding thermodynamic fluxes \cite{Landau10eng,BaymPethick}.
This results in
\begin{equation}
  C^{\kappa}_c = \frac{3}{\pi^2} \int\limits_{-\infty}^{+\infty} d x\, x \Psi^\kappa_c(x)
  f(x) (1-f(x))
\end{equation}
and
\begin{equation}
  C^{\eta}_c = \int\limits_{-\infty}^{+\infty} d x\, \Psi_c^\eta(x)
  f(x) (1-f(x)).
\end{equation}
Unknown functions $\Psi_c(x)$ obey the system of integral
equations, derived by the linearization of the system of transport
equations using anzatz (\ref{eq:F_anzatz}). Without going into
details \cite{BaymPethick,Anderson1987,Shternin2017JPhCS}, the
resulting system of integral equations takes a  form
\begin{widetext}
\begin{equation}
  \Xi (x) f(-x)=\left(1+\frac{x^2}{\pi^2}\right) \Psi_c(x)f(-x) -
  \frac{2}{\pi^2} \int\limits_{-\infty}^{+\infty} d\,x'\,
  f(-x') \frac{x-x'}{1-{\rm e}^{x'-x}} \sum_{i} \lambda_{ci}
  \Psi_i (x'),\label{eq:psi_eq}
\end{equation}
\end{widetext}
where $\Xi(x)=1$ for the shear viscosity and $\Xi(x)=x$ for the
thermal conductivity. This simple form is possible due the
appropriate choice of the  relaxation time $\tau_{c0}$ by
Eq.~(\ref{eq:nu_c0}). All information about the quasiparticle
scattering is encapsulated in the matrix $\lambda_{ci}$ which
depends on the transport coefficient in question. In case of
thermal conductivity, this matrix can be expressed through the
collision frequencies discussed in the previous sections in the
following way:
\begin{equation}
\lambda^\kappa_{ci}=-\frac{5}{4}\frac{{\nu'}^{\kappa}_{ci}}{\nu_{i0}}, \quad i\neq c,
\end{equation}
\begin{equation}
\lambda^\kappa_{cc}=3-\frac{5}{4}\sum\limits_i \frac{\nu^\kappa_{ci}}{\nu_{c0}}-\frac{5}{4}\frac{{\nu'}^{\kappa}_{cc}}{\nu_{c0}}.
\end{equation}
Simplest variational solution  discussed above is
$\Psi^\kappa_c=C^\kappa_c x_c$ and corresponds to the solution of
the linear system $5/4=\sum_i( 3\delta_{ci}-\lambda^\kappa_{ci})
C^\kappa_i$. Once $\lambda^\kappa_{ci}$ is calculated, the system
(\ref{eq:psi_eq}) can be solved numerically and the correction
coefficients $C^\kappa_c$ can be obtained. It turns out, however,
that $\lambda^\kappa_{ci}\approx \delta_{ci}$ in the conditions of
the present study. This is because both $\nu^\kappa_{ci}$ and
$\nu_{c0}$ are dominated by the transverse contribution in its
leading order, while the mixed collision frequencies
${\nu'}_{ci}^{\kappa}$ are of the next order and thus their ratios
to $\nu_{c0}$ are small. As a result, the system of equations
(\ref{eq:psi_eq}) decouples to independent equations for each
species. Therefore the correction to variational solution is given
by the expression for the single-component Fermi liquid with
$\lambda^\kappa =1$. In this case, one obtains
$C^\kappa_c/C^{\kappa,\, \mathrm{Var}}_c=1.2$, see, e.g.,
Ref.~\cite{BaymPethick}. Numerical solution of
Eq.~(\ref{eq:psi_eq}) supports this conclusion, giving
$\kappa_{e\mu}/\kappa_{e\mu}^\mathrm{Var}\approx 1.20-1.22$ in all
considered cases.

The situation is similar for the shear viscosity. The matrix $\lambda^\eta_{ci}$ is
\begin{equation}
\lambda^\eta_{ci}=-\frac{3}{4}\frac{{\nu'}^{\eta}_{ci}}{\nu_{i0}}, \quad i\neq c,
\end{equation}
\begin{equation}
\lambda^\eta_{cc}=1-\frac{3}{4}\sum\limits_i \frac{\nu^\eta_{ci}}{\nu_{c0}}-\frac{3}{4}\frac{{\nu'}^{\eta}_{cc}}{\nu_{c0}},
\end{equation}
and simplest variational result is $\Psi^\eta_c=C^\eta_c$ and
corresponds to the solution of the linear system $1=\sum_i(
\delta_{ci}-\lambda^\eta_{ci}) C^\eta_i$. In this case, since the
collision frequencies for shear viscosity are $q^2$ in the leading
order, in the weak-screening approximation they are much smaller
than $\nu_{c0}$. As a consequence,
$\lambda^\eta_{ci}\approx\delta_{ci}$ as well. Moreover, this
means that variational result does not need to be corrected and
$\eta_{e\mu}=\eta_{e\mu}^\mathrm{Var}$ \cite{BaymPethick}.
Numerical calculations show that this conclusion holds up to 0.1\%
for the present conditions.

\section{Discussion}\label{sec:discuss}

\begin{figure*}[ht]
\begin{minipage}{0.45\textwidth}
\includegraphics[width=\textwidth]{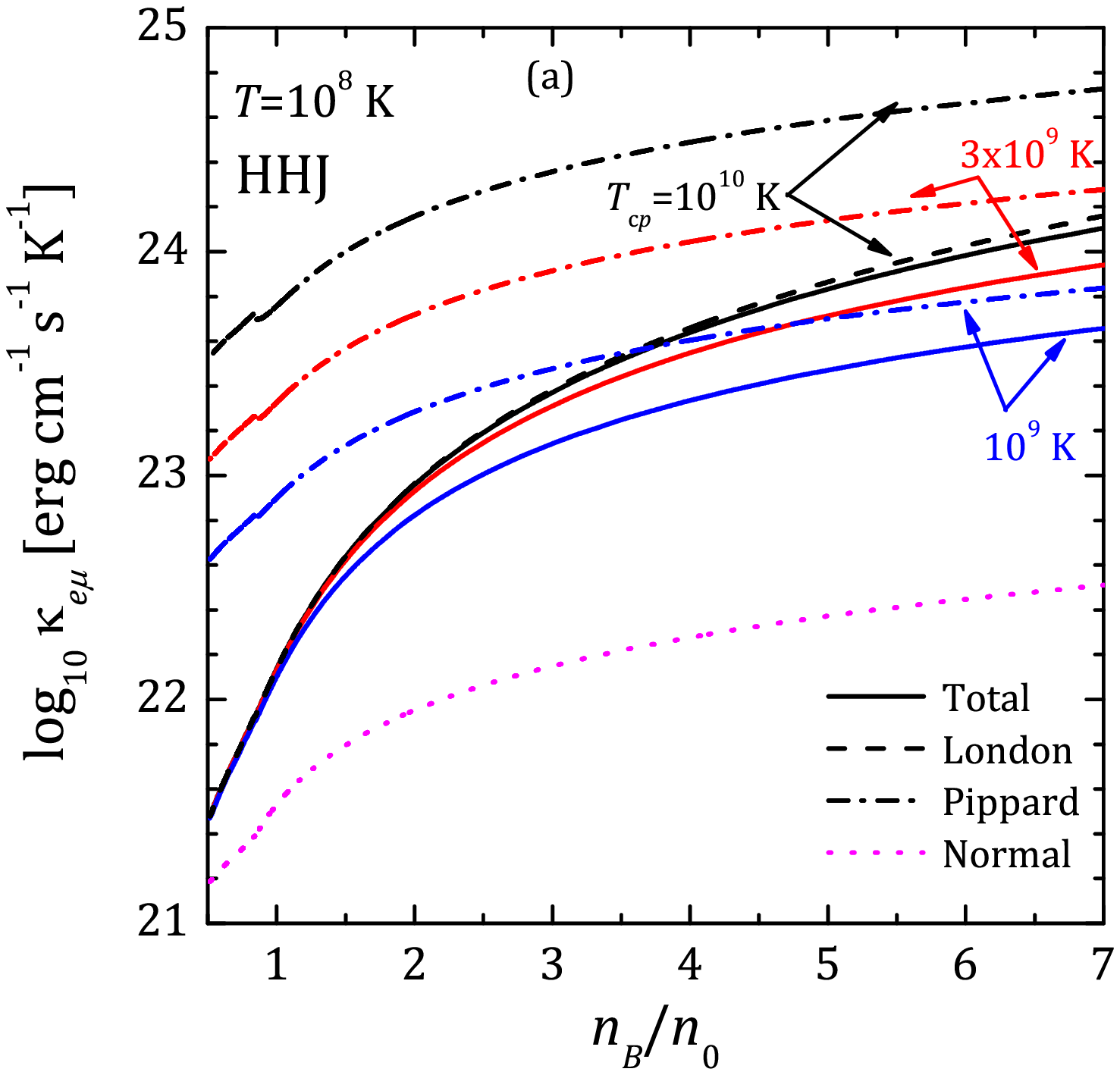}
\end{minipage}
\hspace{0.05\textwidth}
\begin{minipage}{0.45\textwidth}
\includegraphics[width=\textwidth]{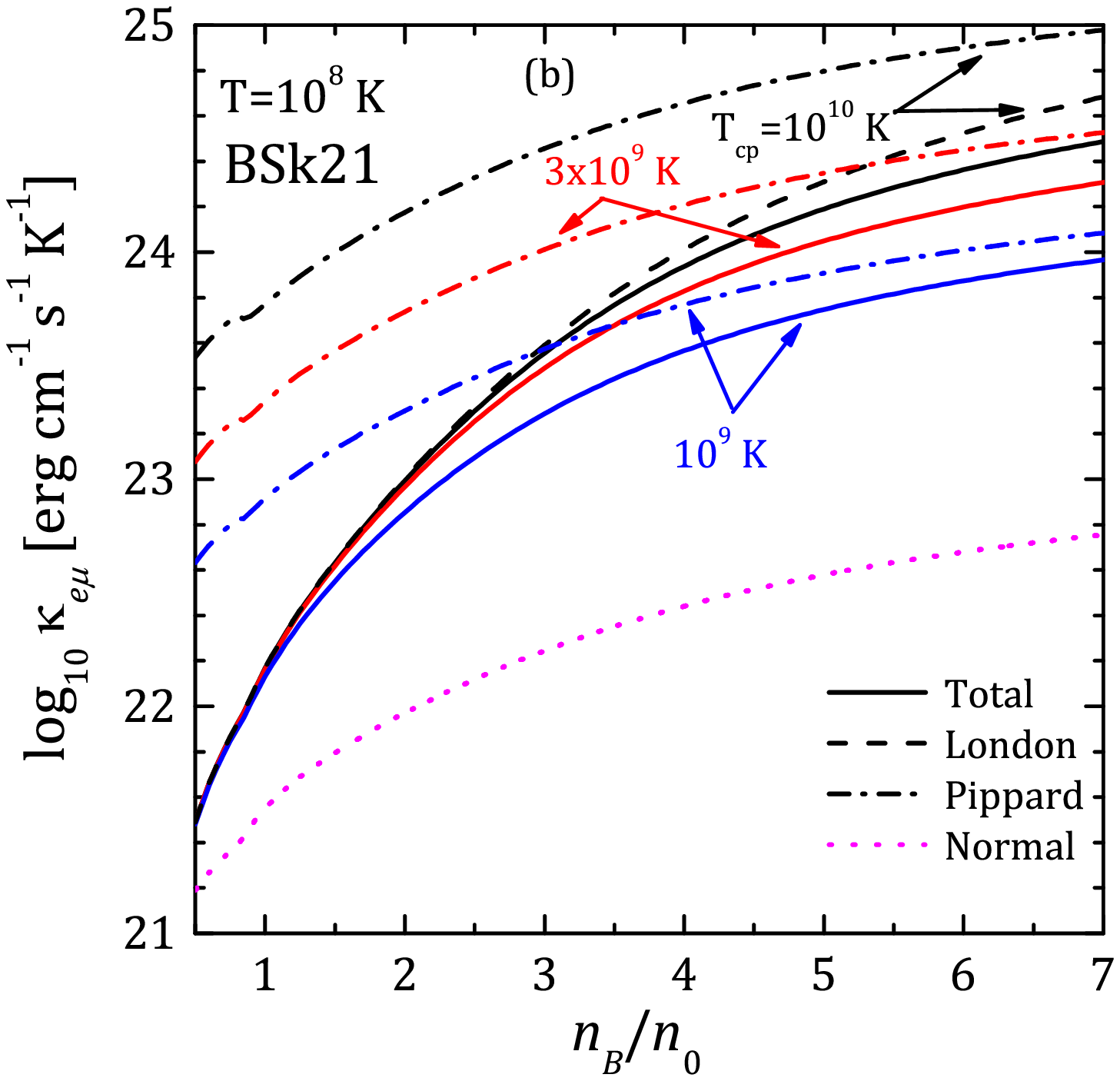}
\end{minipage}
\caption{(Color online). Lepton thermal conductivity
$\kappa_{e\mu}$ as a function of $n_B$  (a) for the  HHJ EOS and
(b) for the  BSK21 EOS, $T=10^8$~K, and for three values of
$T_{\mathrm{c}p}=10^{10}$, $3\times 10^9$, and $10^9$~K as
indicated in the plot. Solid lines show the results of the present
paper, dash-dotted lines correspond to the Pippard limit, and
dashed lines to the London limit of transverse screening. Dotted
lines show the calculations for normal (not superconducting)
matter.  }\label{fig:kappa}
\end{figure*}

The results of the previous sections can be used for calculating
the lepton contribution to transport coefficients of
superconducting NS cores at not-too-high temperatures. Since these
coefficients are governed by the electromagnetic interactions, the
obtained results are applicable for any npe$\mu$ EOS of a neutron
star core, provided the particle fractions and proton effective
masses are known. It is clear from examining
Eqs.~(\ref{eq:kappa_emu_Meiss})--(\ref{eq:eta_emu_Pipp}), that the
increase in proton fraction at a  given baryon density lead to
increase in both $\kappa_{e\mu}$ and $\eta_{e\mu}$. At the same
time, the parameter $A$ increases with $x_p$ as well, so the
crossover between the London and Pippard regimes occurs at lower
$n_B$ for the EOS with higher proton fraction. Specifically,
results are illustrated for two EOSs, HHJ and BSk21, and as
before, the proton effective mass $m^*_p=0.8 m_u$ is used. (The
effect of effective mass variation is discussed separately below.)
For completeness, all results discussed in this section employ
exact solution of the system of transport equations taking into
account all kinematical corrections as discussed in
Secs.~\ref{sec:collfreq_corr}--\ref{sec:corr_Var}.

Figure~\ref{fig:kappa} shows the total lepton thermal conductivity
$\kappa_{e\mu}$ as a function of $n_B$ for the HHJ EOS [panel (a)]
and the BSk21 EOS [panel (b)], $T=10^8$~K and three values of
$T_{\mathrm{c}p}=10^{10}$~K, $3\times 10^9$~K, and $10^9$~K.
Remember, that the lepton thermal conductivity in a
superconducting NS core scales as $\kappa_{e\mu}\propto T^{-1}$ as
in the normal Fermi liquid. Solid lines give the results of the
present paper with the corrected description of the transverse
screening. Dash-dotted lines are calculated taking the screening
in the Pippard limit, as in Ref.~\cite{ShterninYakovlev2007PhRvD}.
According to Eq.~(\ref{eq:kappa_emu_Pipp}), in this limit
$\kappa_{e\mu}$ is approximately proportional to $\Delta$.
Clearly, these results strongly overestimate $\kappa_{e\mu}$,
especially at lower densities and higher values of
$T_{\mathrm{c}p}$, which correspond to low values of the $A$
parameter. The dashed lines in Fig.~\ref{fig:kappa} show
$\kappa_{e\mu}$ calculated employing  the transverse screening in
the London limit. In this limit, $\kappa_{e\mu}$ is independent of
$\Delta$, see Eq.~(\ref{eq:kappa_emu_Pipp}), so only the single
dashed line is present in Fig.~\ref{fig:kappa}. Thermal
conductivity calculated in this limit also overestimates
$\kappa_{e\mu}$. For low densities and/or high $T_{\mathrm{c}p}$,
this overestimation is small. For instance, for
$T_{\mathrm{c}p}=10^{10}$~K, dashed lines give rather good
approximation for $\kappa_{e\mu}$ (compare with the solid lines)
for all shown densities and for both EOSs. In contrast, for high
densities and  $T_{\mathrm{c}p}=10^9$~K, the Pippard limit gives
much better approximation than the London one. Comparing left and
right panels of Fig.~\ref{fig:kappa}, one can see that the
difference between $\kappa_{e\mu}$ calculated in the Pippard limit
and the correct value is smaller for the BSk21 EOS than for the
HHJ EOS. Similarly, the difference between the London-limit
calculations and exact ones (solid lines) is larger for the BSk21
EOS than for the HHJ EOS. This is a consequence of the larger
value of the $A$ parameter for the BSk21 EOS (see
Fig.~\ref{fig:A}). For comparison, with dotted lines in
Fig.~\ref{fig:kappa}, $\kappa_{e\mu}^{N}$ calculated for the
non-superconducting case is plotted. Again, all terms in the
interaction are included, although the leading-order
Eq.~(\ref{eq:kappa_emu_trans_norm}) gives a good approximation
(notice, that correction to the variational solution is negligible
in this case \cite{ShterninYakovlev2007PhRvD}). Since
$\kappa_{e\mu}^N$ in leading order does not depend on temperature
(see Sec.~\ref{sec:kin_form}), the difference between
$\kappa_{e\mu}$  and $\kappa_{e\mu}^N$ increases with lowering $T$
\cite{ShterninYakovlev2007PhRvD}. Taking this in mind and looking
at the lower-density region in the left panel of
Fig.~\ref{fig:kappa} one can naively suggest that with increase of
temperature, the normal-matter $\kappa_{e\mu}^{N}$ would become
larger than $\kappa_{e\mu}$ for superconducting matter. This is
not so, since in these conditions the assumption of the dominance
of the proton contribution to the transverse screening will break
down. In this case, one needs to include the dynamical
contribution from the normal constituents of matter (leptons) to
the transverse screening, see below. Presence of the normal matter
contribution to screening effectively limits the collision
frequencies from above,  making them lower than in the normal
case.

\begin{figure*}[ht]
\begin{minipage}{0.45\textwidth}
\includegraphics[width=\textwidth]{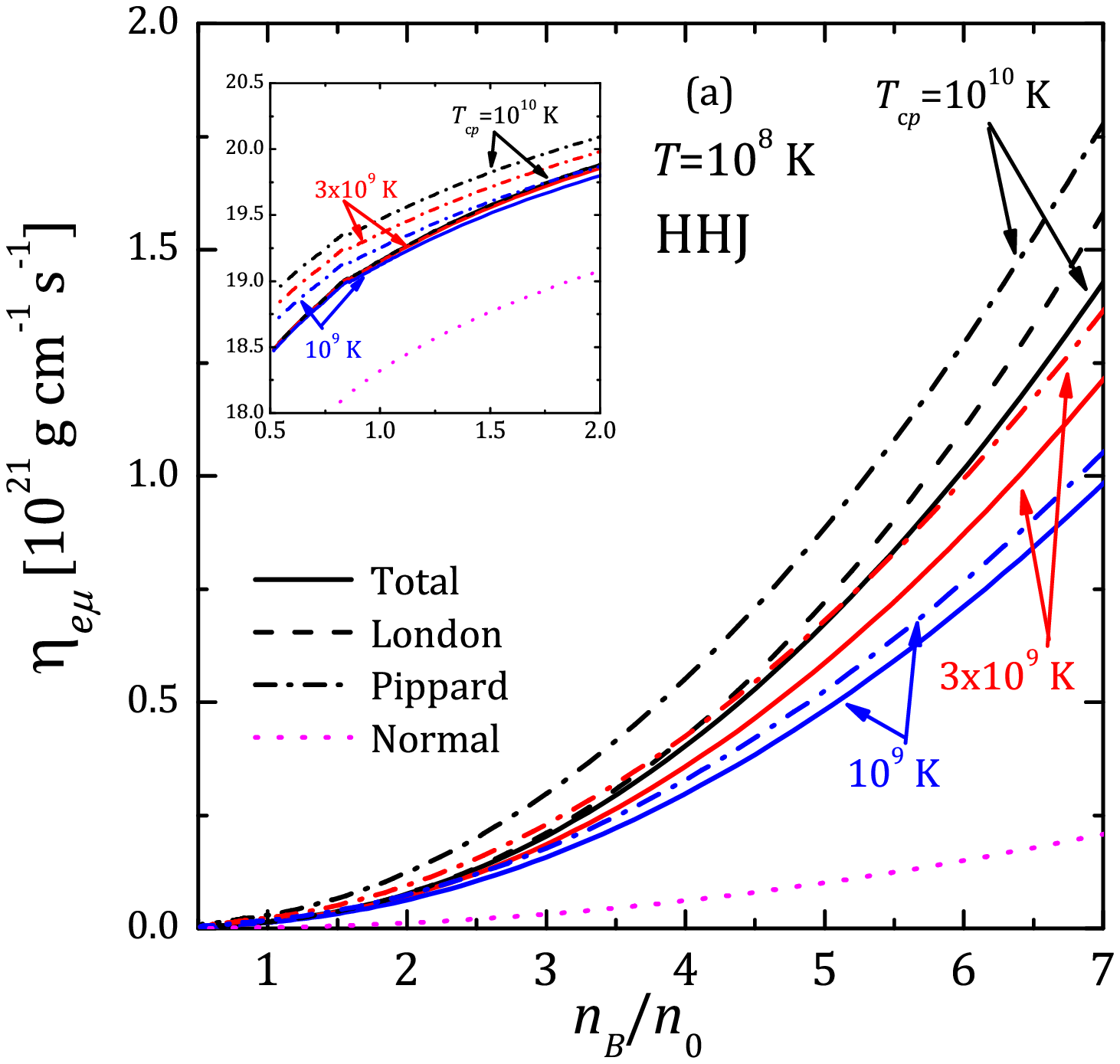}
\end{minipage}
\hspace{0.05\textwidth}
\begin{minipage}{0.45\textwidth}
\includegraphics[width=\textwidth]{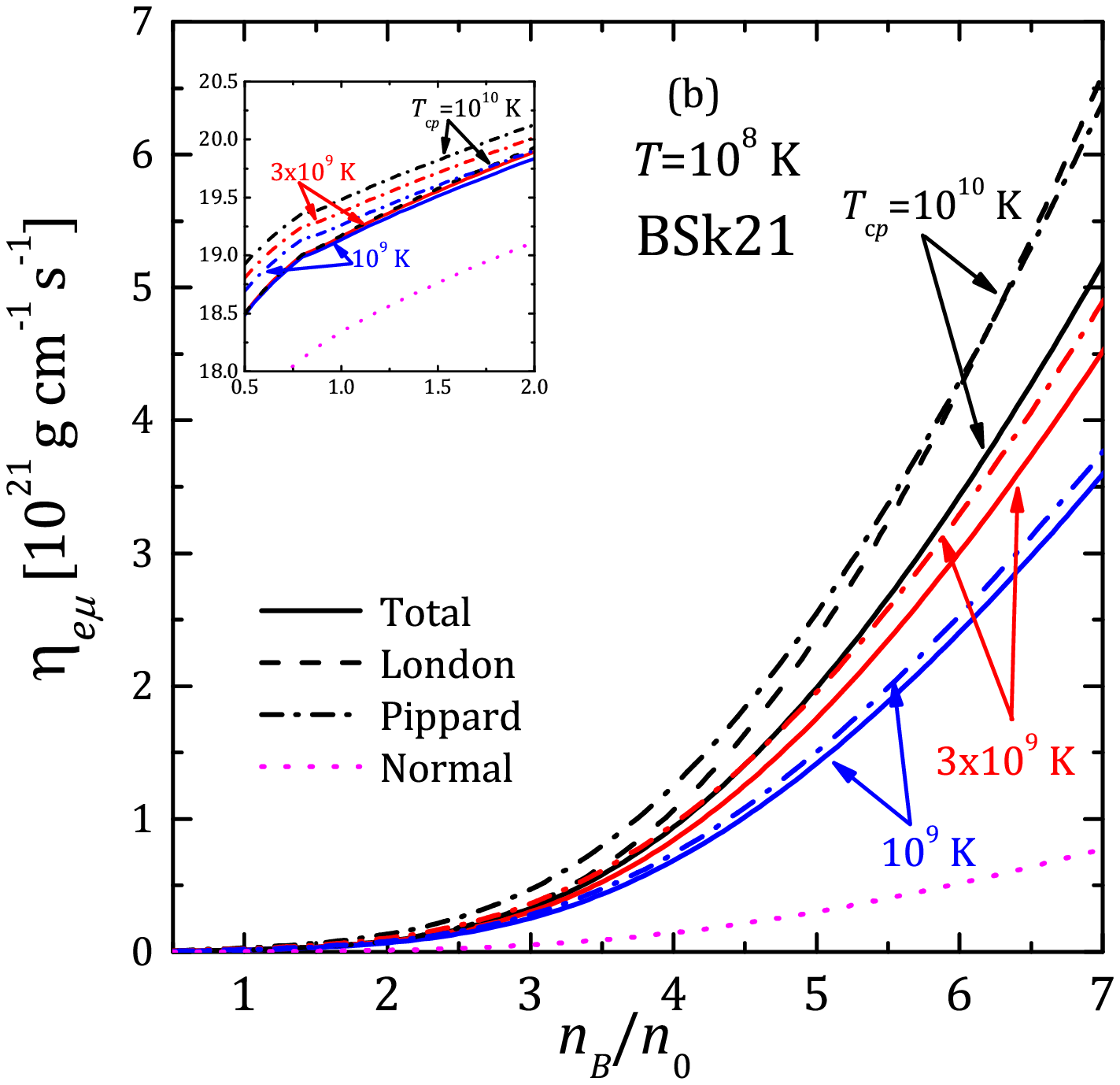}
\end{minipage}
\caption{(Color online). Lepton shear viscosity  (a) for the HHJ
EOS and (b) for the BSK21 EOS as a function of $n_B$. The
parameters of calculations  and notations are the same as in
Fig.~\ref{fig:kappa}. Notice, that the linear scale is used in
this plot. Insets enlarge the region of $n_B$ up to $2n_0$, where
the logarithmic scale for $\eta_{e\mu}$ is used.}\label{fig:eta}
\end{figure*}

Similar calculations for the shear viscosity $\eta_{e\mu}$ are
shown in Fig.~\ref{fig:eta}. Qualitatively, the situation is the
same as for the thermal conductivity, although the differences
between results in various approximations are less dramatic. This
is a consequence of, first, the weaker dependence of the
transverse collision frequencies $\nu^\eta_{ci}$ on the screening
than found for  $\nu^\kappa_{ci}$ and, second, of the larger
contribution of the longitudinal part of the interaction to
$\eta_{e\mu}$ than to $\kappa_{e\mu}$. For instance, the older
results of Ref.~\cite{ShterninYakovlev2008PhRvD} calculated in the
Pippard limit (dash-dotted lines in Fig.~\ref{fig:eta}) are
acceptable for $T_{\mathrm{c}p}<3\times 10^9$~K, especially at
higher densities. At most, the use of the Pippard limit results in
an overestimation of $\eta_{e\mu}$ by a factor of 2.5 for
$T=10^9$~K and lowest densities. This is not seen in
Fig.~\ref{fig:eta} because of the linear scale, and is illustrated
in the insets that show $\eta_{e\mu}$ up to $n_B=2n_0$ with the
logarithmic scale. In the inset plots, due to a low density, the
difference between $\eta_{e\mu}$ calculated for various
$T_{\mathrm{c}p}$ and those calculated in London limit is barely
seen. On the over hand, the overestimation that results from using
the Pippard expression becomes visible.  $\eta_{e\mu}^N$
calculated for non-superconducting matter is shown in
Fig.~\ref{fig:eta} with dotted lines. Notice again, that the
relation $\eta_{e\mu}^N\propto T^{-5/3}$ given by
Eq.~(\ref{eq:eta_emu_trans_norm}) works well only at low
temperatures, where the transverse part of the interaction starts
to dominate \cite{ShterninYakovlev2008PhRvD}. Both the shear
viscosity and the thermal conductivity  for the BSk21 EOS are
larger than those for the HHJ EOS. This is a consequence of
different particle number fractions in these models.

\begin{figure*}[ht]
\begin{minipage}{0.45\textwidth}
\includegraphics[width=\textwidth]{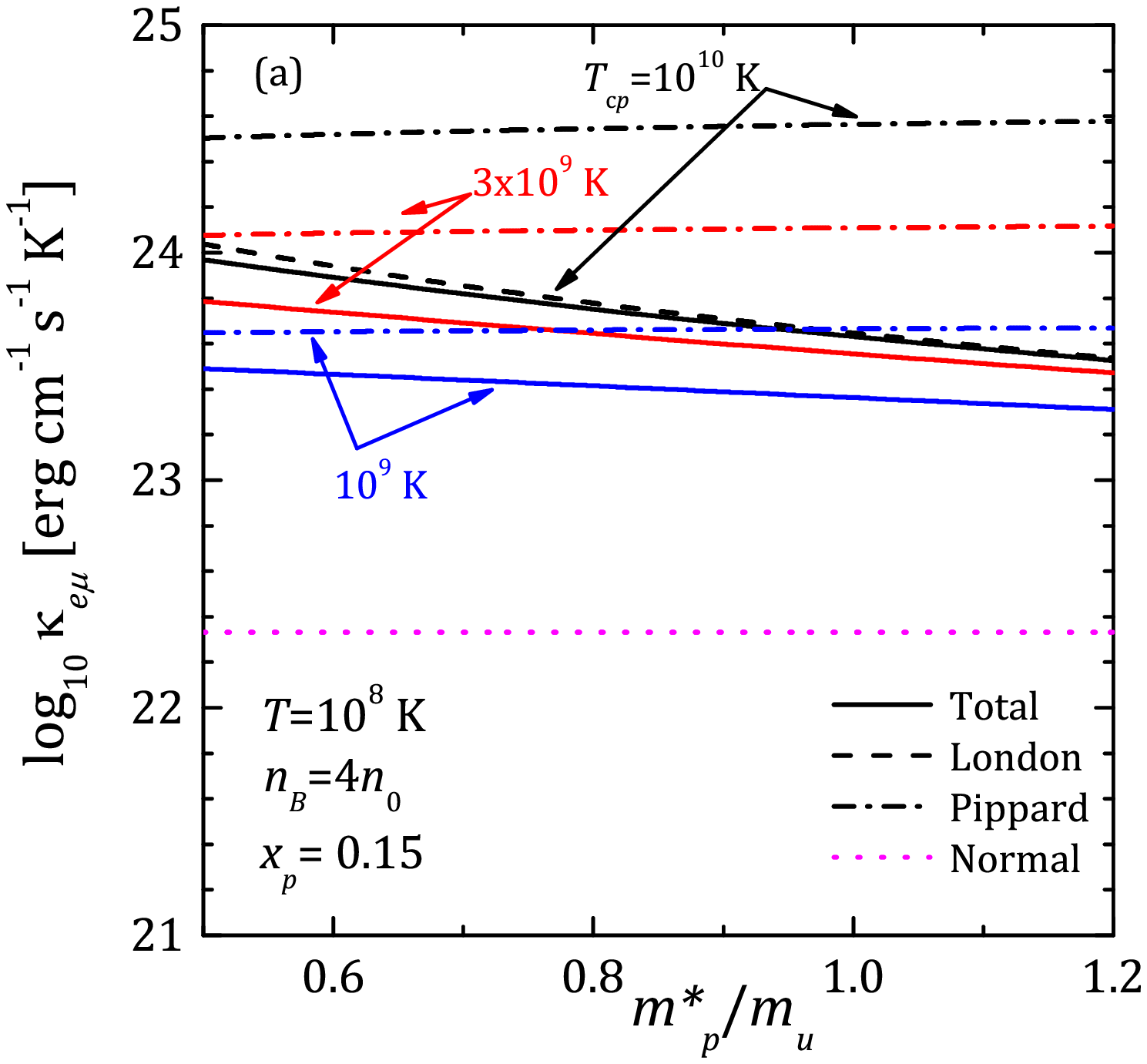}
\end{minipage}
\hspace{0.05\textwidth}
\begin{minipage}{0.45\textwidth}
\includegraphics[width=\textwidth]{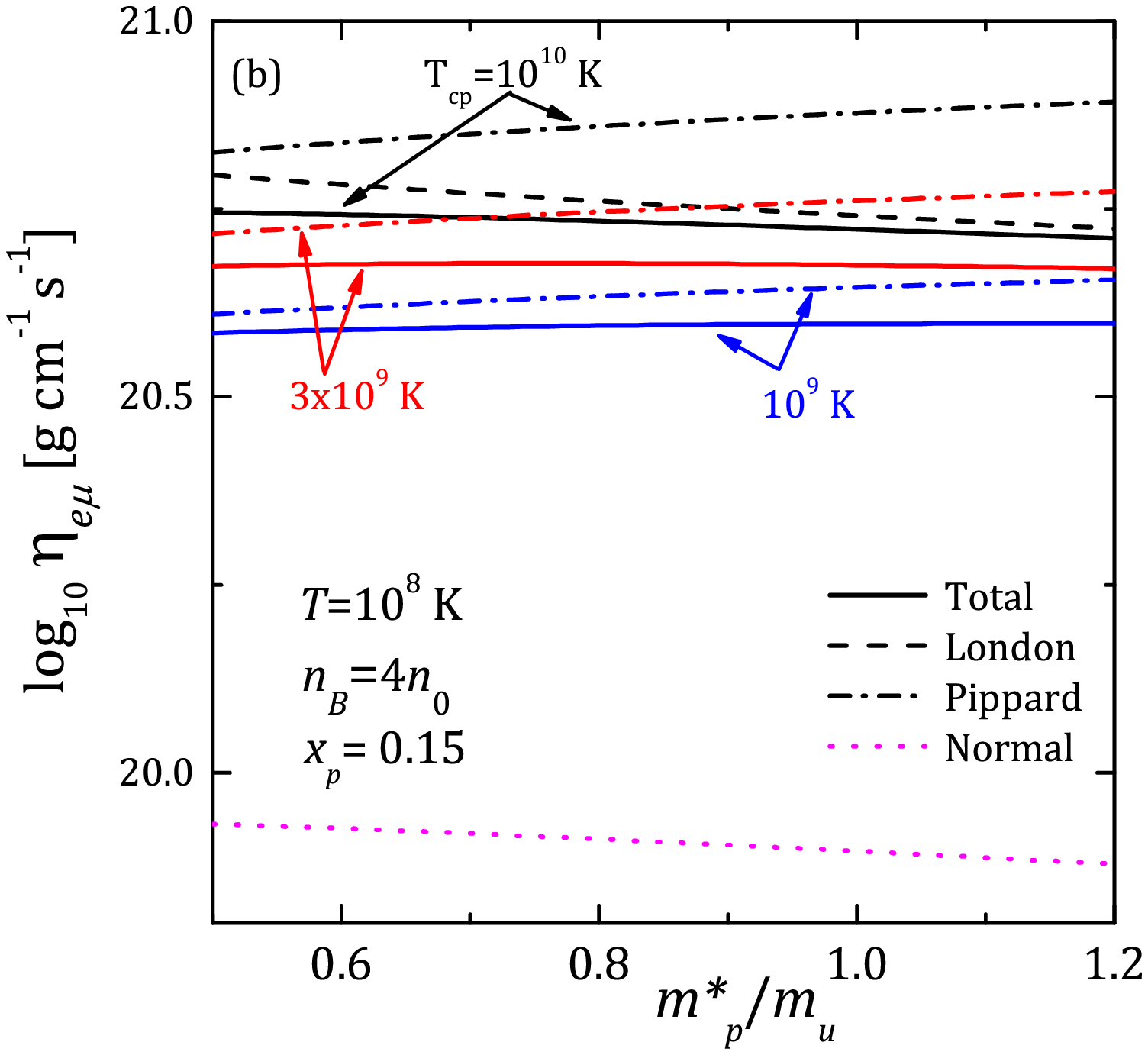}
\end{minipage}
\caption{(Color online). Lepton thermal conductivity  (a) and
shear viscosity (b) calculated for $T=10^8$~K, $n_B=4n_0$, and
proton fraction $x_p=0.15$ as functions of the proton effective
mass $m^*_p$.  The notations are the same as in
Fig.~\ref{fig:kappa}.}\label{fig:kin_meff}
\end{figure*}

In the previous discussion the constant (density-independent)
proton effective mass was employed. It is instructive to look how
the results depend on $m_p^*$. This is illustrated in
Fig.~\ref{fig:kin_meff} where the thermal conductivity [panel (a)]
and shear viscosity [panel (b)] are plotted as a function of
$m^*_p$ for the density $n_B=4n_0$ and proton fraction $x_p=0.15$.
These are some typical values  and do not correspond to a specific
EOS. The line types and notations are similar to those in
Figs.~\ref{fig:kappa}-\ref{fig:eta}. Asymptotic expression in the
Pippard limit in the leading order,
Eqs.~(\ref{eq:kappa_emu_Pipp})--(\ref{eq:eta_emu_Pipp}) do not
depend on the proton effective mass. Some dependence on $m^*_p$
demonstrated by the dash-dotted lines in Fig.~\ref{fig:kin_meff}
is due to the corrections beyond the leading order. This
dependence is more pronounced for the shear viscosity than for the
thermal conductivity, in accordance with the discussion in
Sec.~\ref{sec:collfreq_corr}. Similar arguments apply for the
transport coefficients of the normal matter shown with dotted
lines in Fig.~\ref{fig:kin_meff}. In contrast, the asymptotic
expressions in the London limit,
Eqs.~(\ref{eq:kappa_emu_Meiss})--(\ref{eq:eta_emu_Meiss})
explicitly depend on the proton effective mass through the
Meissner momentum $q_M$. According to Eq.~(\ref{eq:Meissner}), an
increase in $m^*_p$ leads to decrease in $q_M$ and hence to
decrease of $\kappa_{e\mu}$ and $\eta_{e\mu}$. As seen in
Fig.~\ref{fig:kin_meff}, this decrease is larger for
$\kappa_{e\mu}$ than for $\eta_{e\mu}$ because of the weaker
$q_M$-dependence of the latter, see
Eqs.~(\ref{eq:kappa_emu_Meiss})--(\ref{eq:eta_emu_Meiss}). The
dependence of the results of the full calculations on the
effective mass is in between the discussed limiting cases. Since
the parameter $A$ decreases with $m_p^*$ [see
Eq.~(\ref{eq:A_natural})], London limiting expressions work better
for larger $m_p^*$. Figure~\ref{fig:kin_meff} shows that the use
of the variable proton effective mass instead of the constant
$m_p^*$ approximation can change the results illustrated in
Figs.~\ref{fig:kappa}-\ref{fig:eta} quantitatively, but not
qualitatively. In principle one should use the density dependence
of $m_p^*$ consistent with the chosen EOS, but this is rarely
provided.

\begin{figure*}[ht]
\begin{minipage}{0.45\textwidth}
\includegraphics[width=\textwidth]{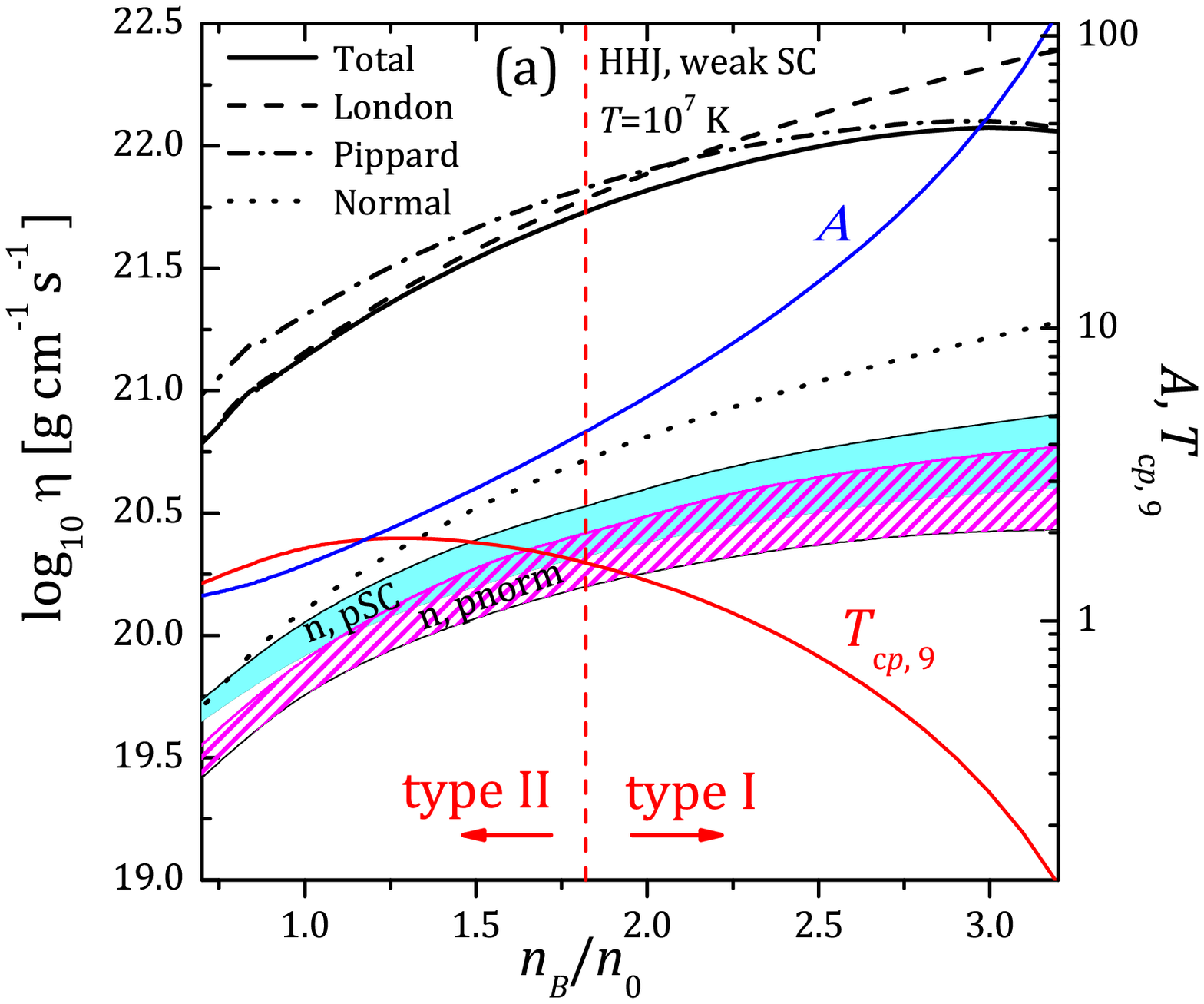}
\end{minipage}
\hspace{0.05\textwidth}
\begin{minipage}{0.45\textwidth}
\includegraphics[width=\textwidth]{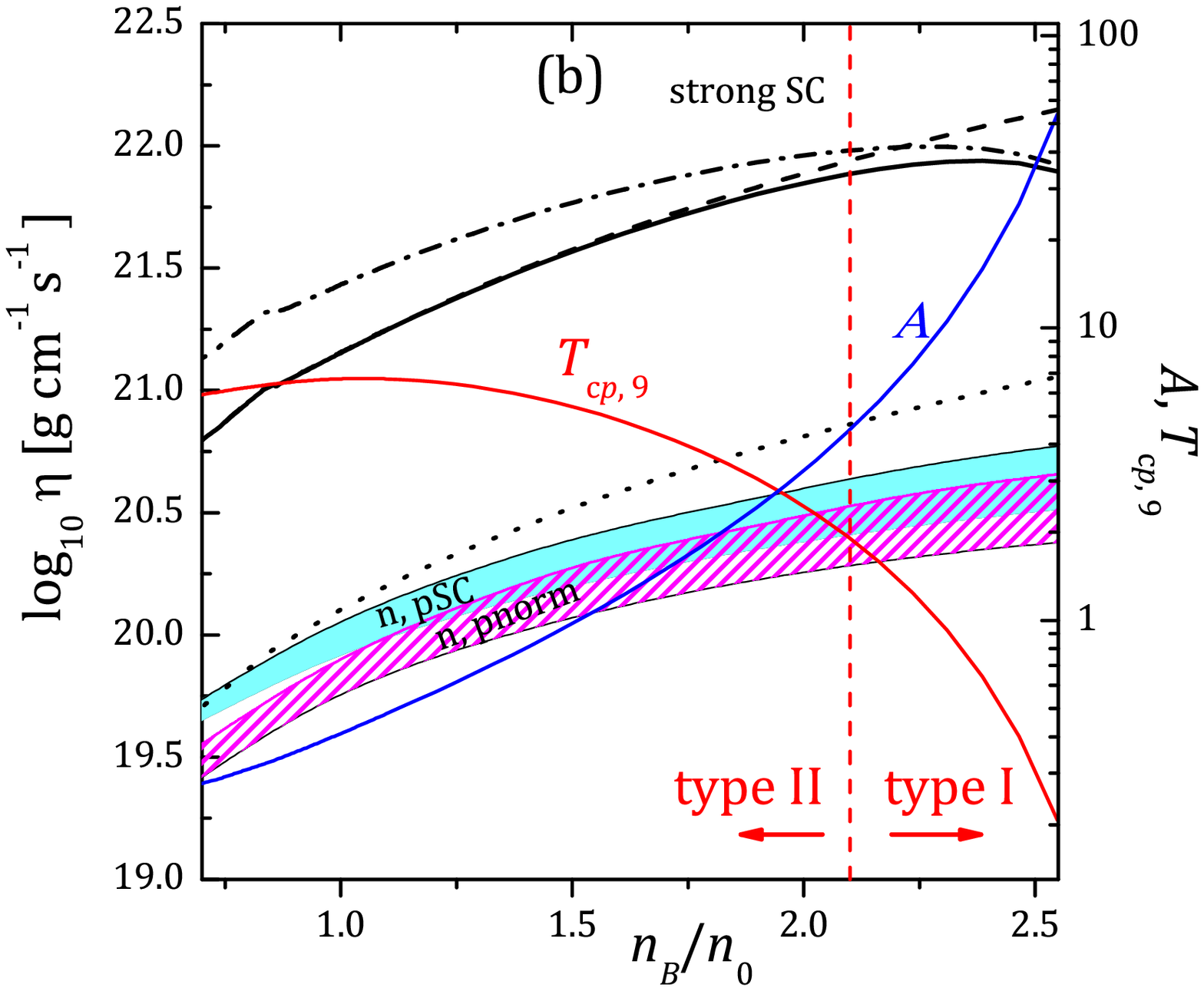}
\end{minipage}
\begin{minipage}{0.45\textwidth}
\includegraphics[width=\textwidth]{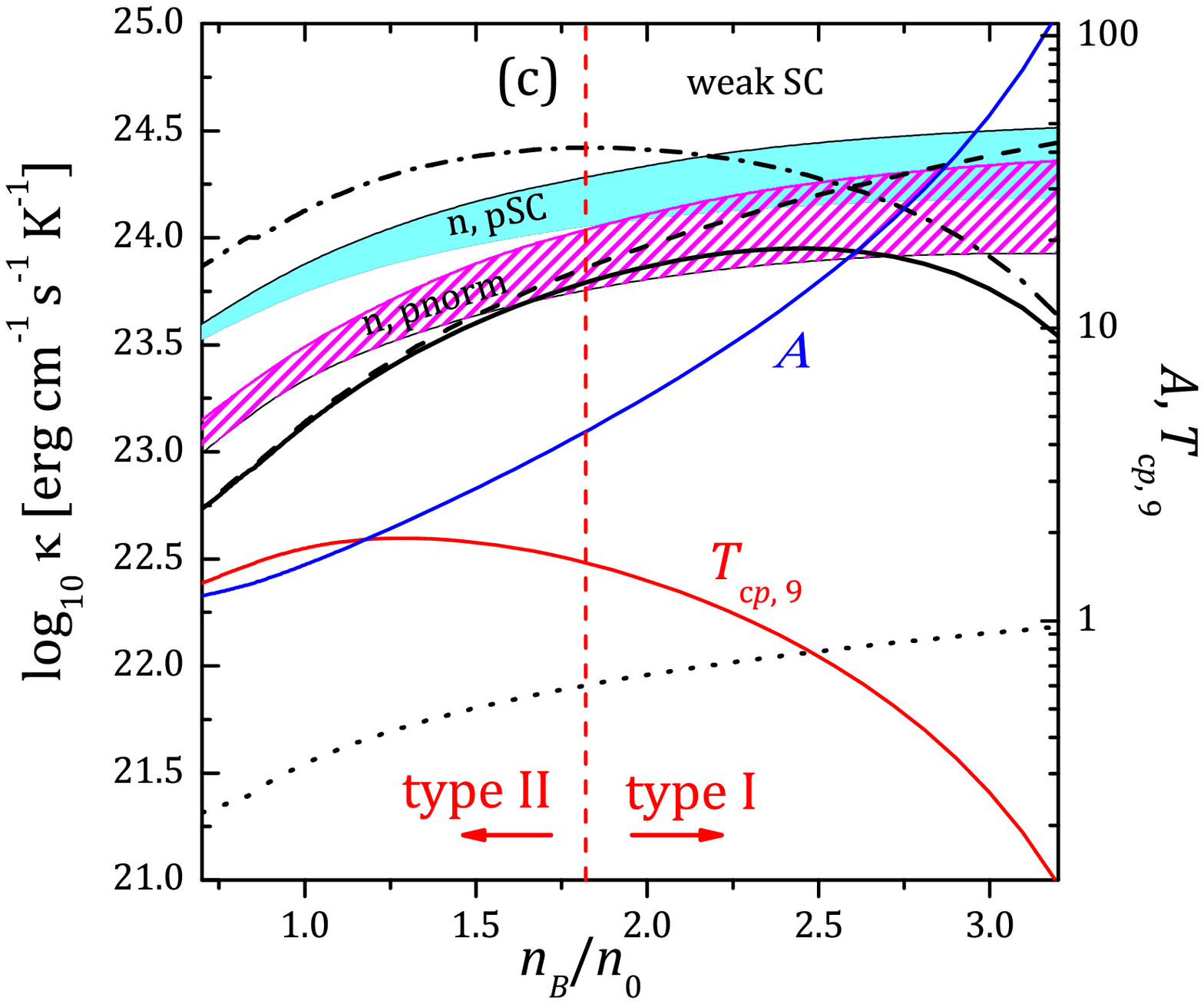}
\end{minipage}
\hspace{0.05\textwidth}
\begin{minipage}{0.45\textwidth}
\includegraphics[width=\textwidth]{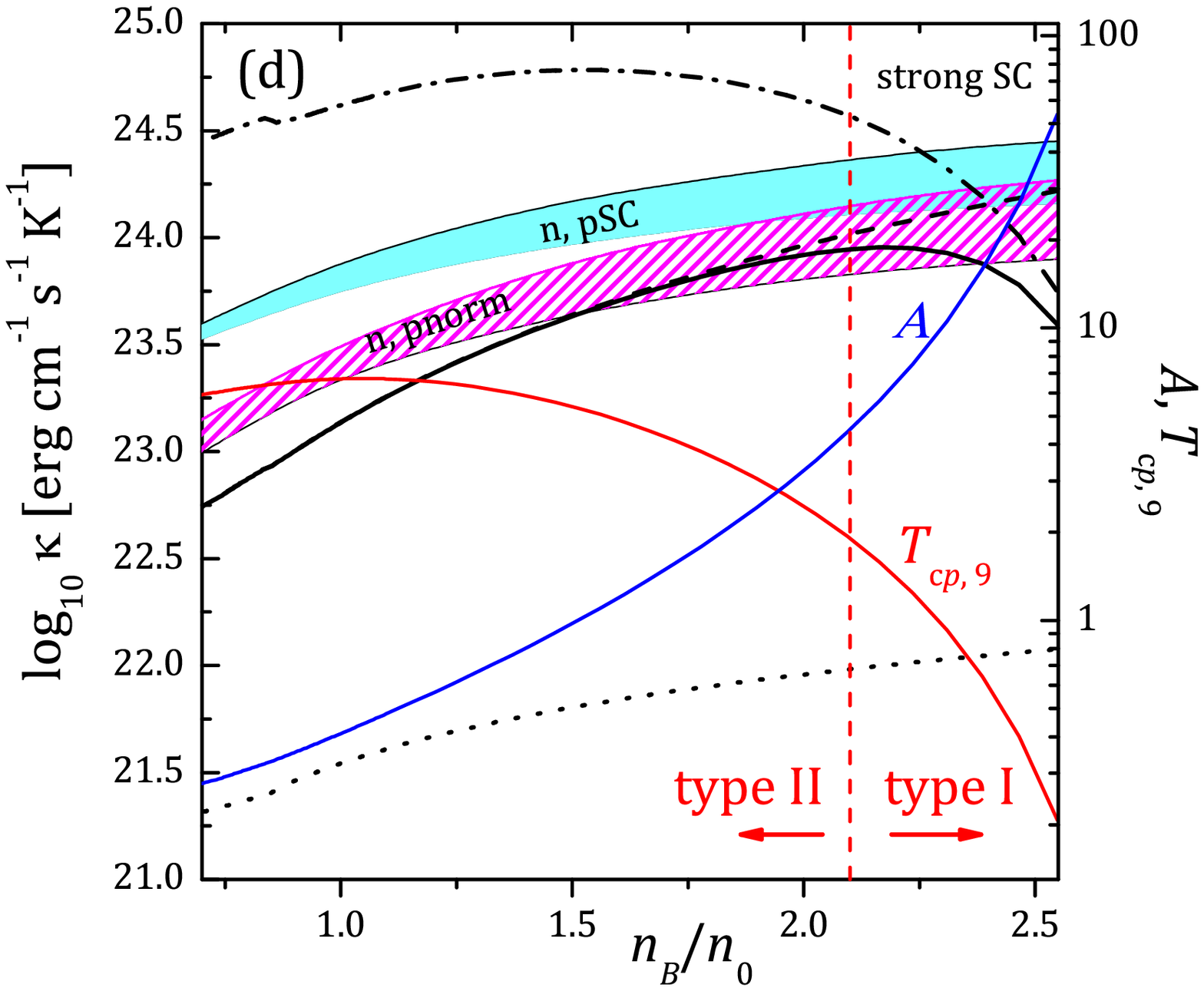}
\end{minipage}
\caption{(Color online). Shear viscosity (a)--(b) and thermal
conductivity (c)--(d) in the HHJ NS core for weak proton
superconductivity model [panels (a) and (c)] and strong
superconductivity model [panels (b) and (d)] discussed in the
text. Temperature is set to $T=10^7$~K. As in previous figures,
solid lines give full results of the present work, dash-dotted
lines correspond to calculations in the Pippard limit, and dashed
lines -- to calculations in London limit. Dotted lines show the
transports coefficients in normal matter. Lower hatched strips
marked `n, pnorm' and upper filled strips marked `n, pSC' give the
uncertainty bands for neutron transport coefficients in normal and
superconducting matter, respectively. Thin solid lines and the
right scales in each panel show the values of
$T_{\mathrm{c}p,9}=T_{\mathrm{c}p}/10^9~{\rm K}$ and $A$ for each
model. Vertical dashed lines ($A=\sqrt{2}\pi$) divide type-II and
type-I superconductivity regions. See text for
details}\label{fig:kappa-eta-And-APR}
\end{figure*}

\begin{figure*}[ht]
\begin{minipage}{0.45\textwidth}
\includegraphics[width=\textwidth]{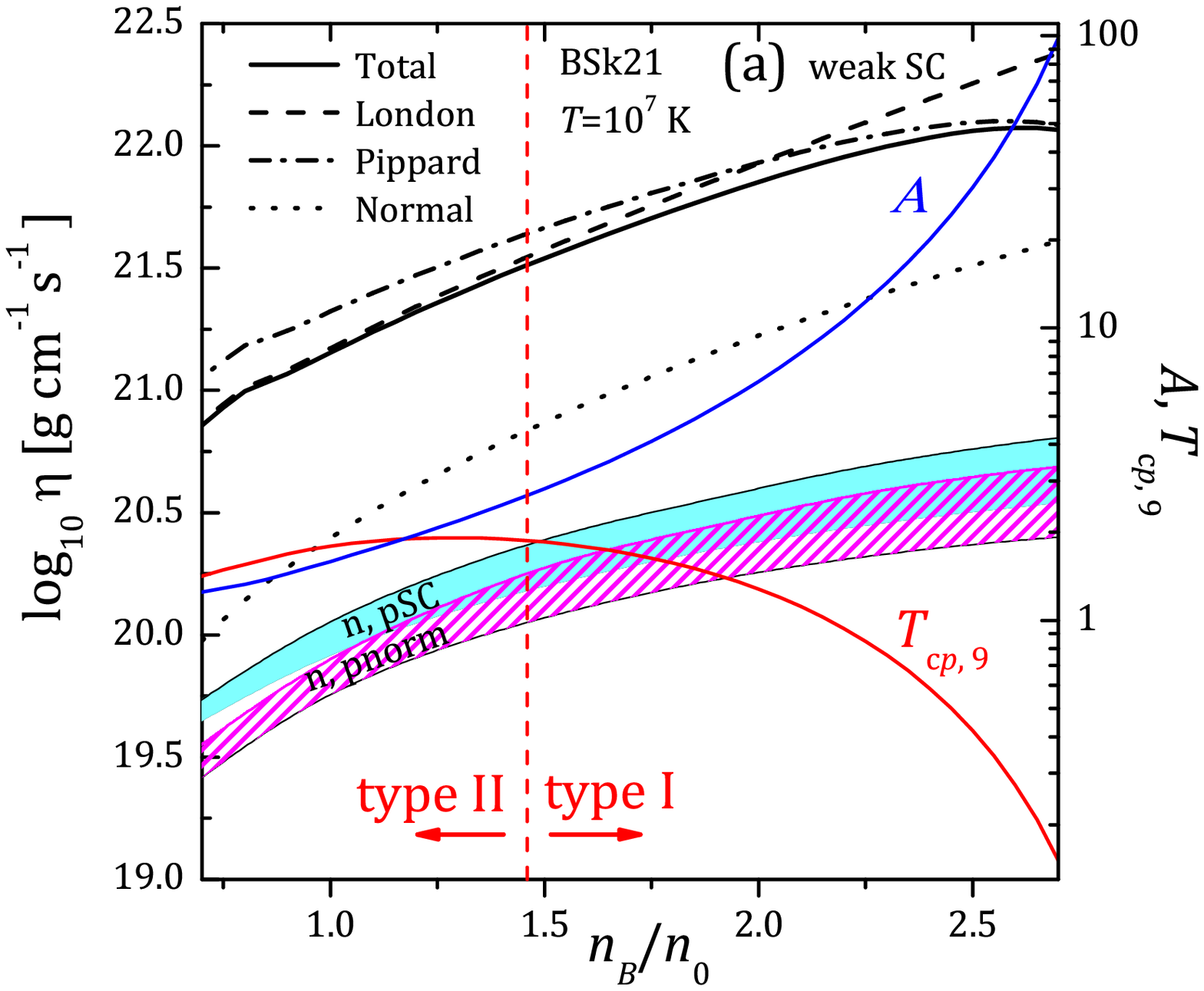}
\end{minipage}
\hspace{0.05\textwidth}
\begin{minipage}{0.45\textwidth}
\includegraphics[width=\textwidth]{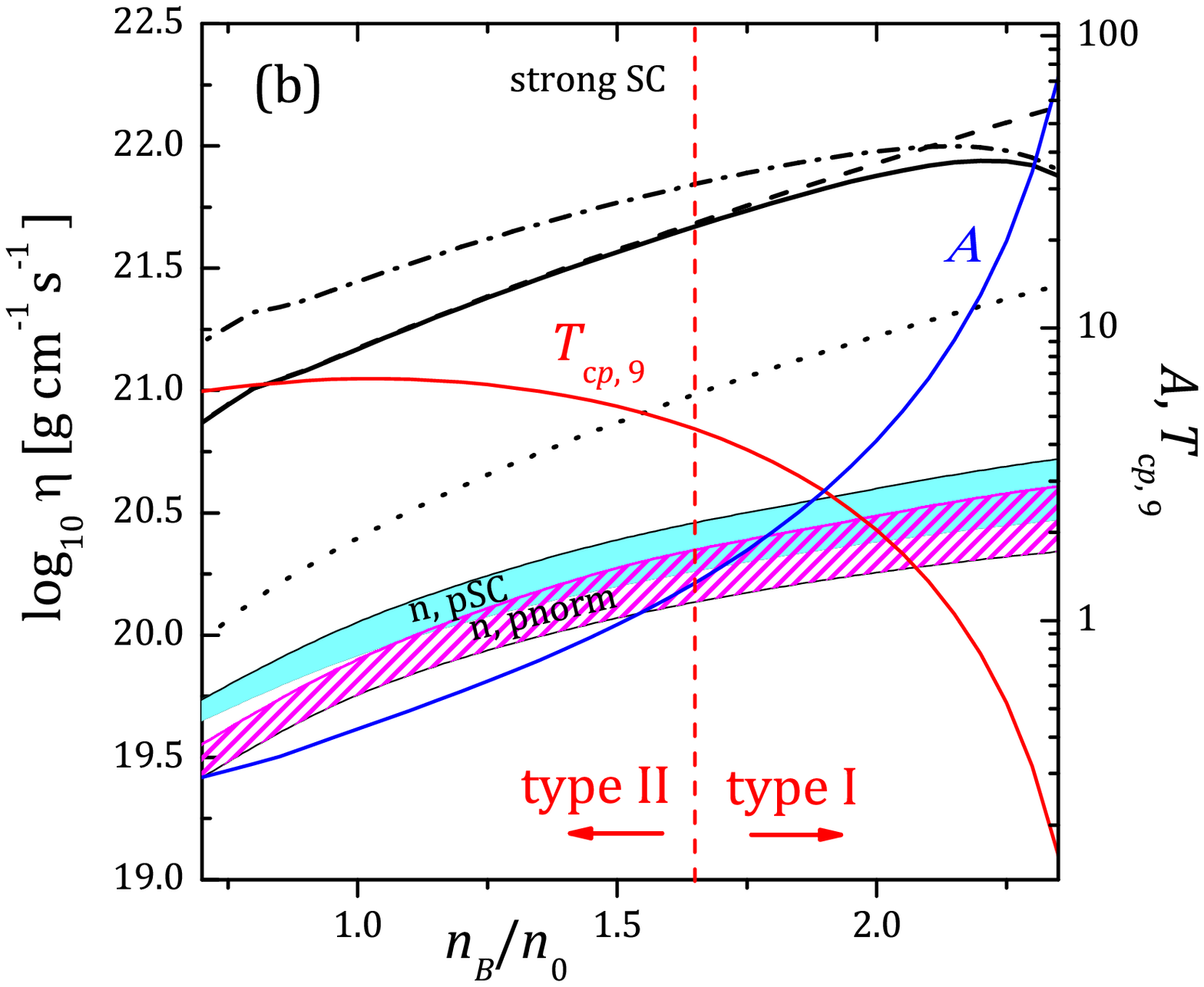}
\end{minipage}
\begin{minipage}{0.45\textwidth}
\includegraphics[width=\textwidth]{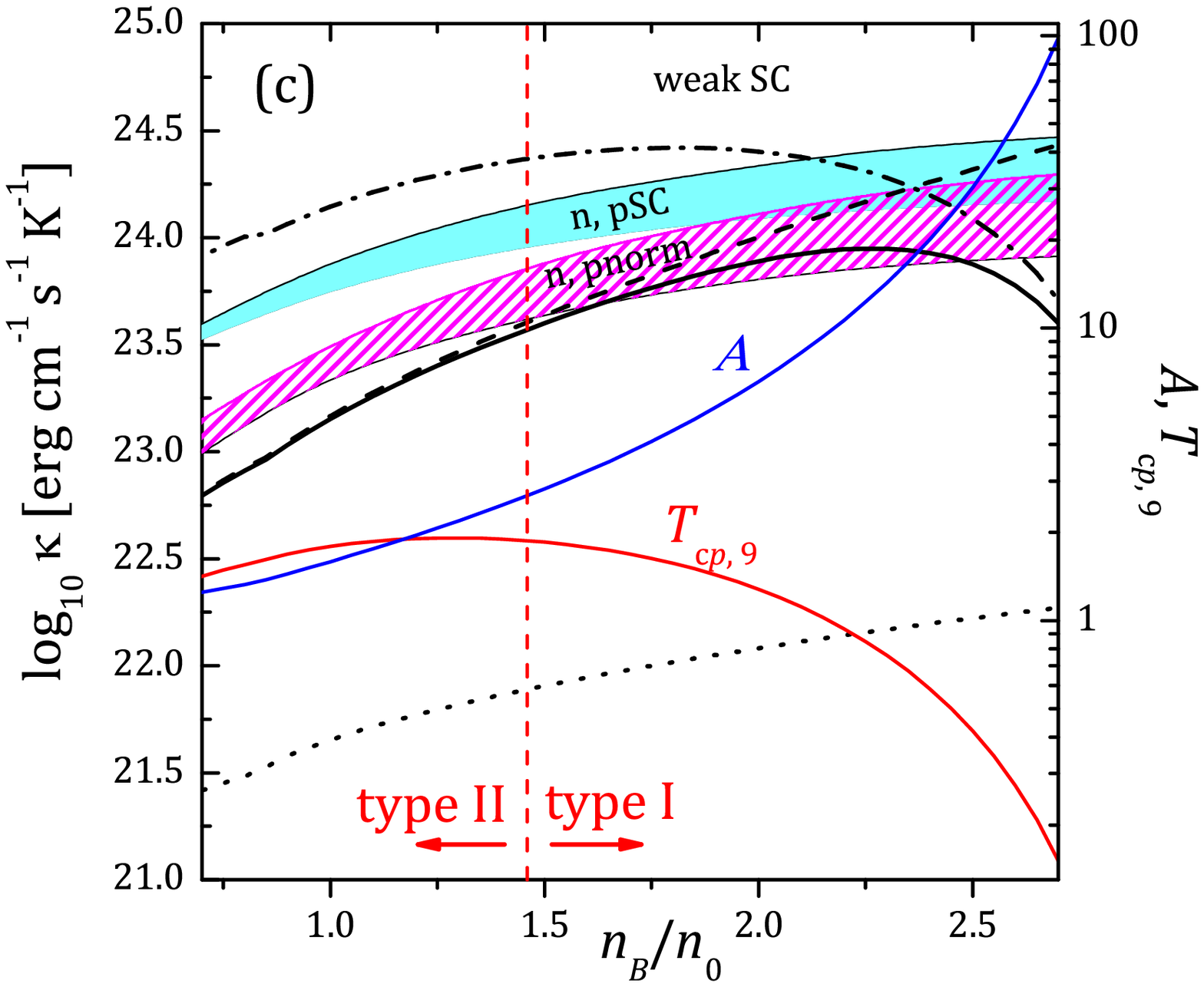}
\end{minipage}
\hspace{0.05\textwidth}
\begin{minipage}{0.45\textwidth}
\includegraphics[width=\textwidth]{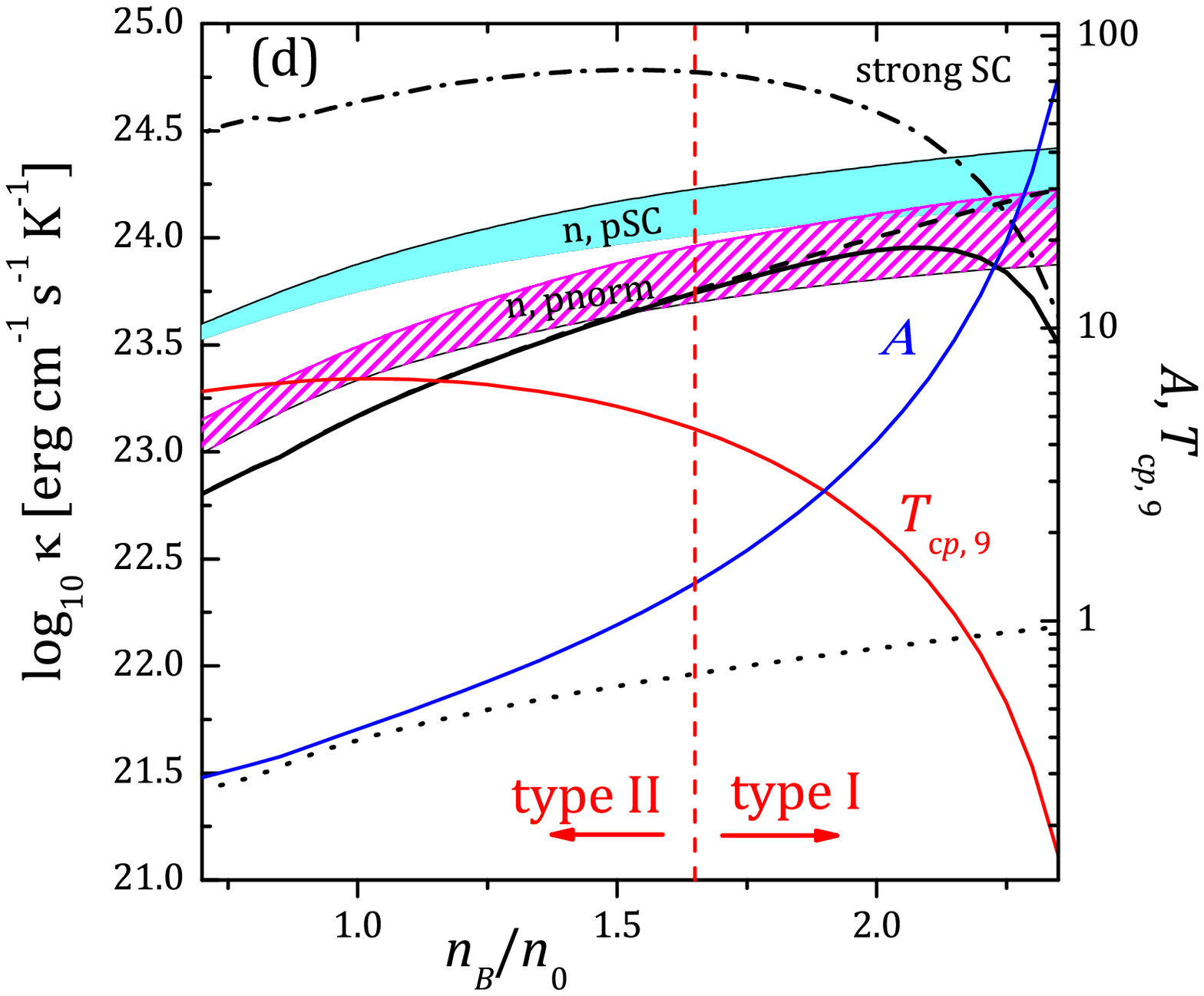}
\end{minipage}
\caption{(Color online). Same as Fig.~\ref{fig:kappa-eta-And-APR}
but for the BSk21 EOS. Notations are the same as in
Fig.~\ref{fig:kappa-eta-And-APR}.}\label{fig:kappa-eta-And-BSk}
\end{figure*}

The proton critical temperature in the NS core is not constant but
is actually density-dependent. It is instructive to illustrate the
results by considering `realistic' profiles $T_{\mathrm{c}p}(n_B)$
in the NS core.This is done in Figs.~\ref{fig:kappa-eta-And-APR}
and \ref{fig:kappa-eta-And-BSk} for the HHJ and the BSK21 EOSs,
respectively.\footnote{Note that again $m^*_p=0.8m_u$.} There
exists a variety of calculations of the critical density profiles
in a literature, each of which is based on a specific microscopic
model. The results of these calculations generally do not agree
with each other. In these circumstances it is instructive to rely
on the phenomenological profiles $T_{\mathrm{c}p}(n_B)$ instead of
trying to handle EOS and superconductivity properties
self-consistently \cite{Kaminker2001A&A,Anderson2005NuPhA}. It is,
however, reasonable to make these phenomenological models to
resemble the extreme cases available on the market. Taking this in
mind, following \citet{Glampedakis2011MNRAS}, I take two profiles
of $T_{\mathrm{c}p}(n_B)$ denoted by `e' and `f' in
Ref.~\cite{Anderson2005NuPhA}. These models are constructed by
applying the phenomenological parametrization suggested by
\citet{Kaminker2001A&A} to the results of microscopic calculations
of the proton $^1$S$_0$ gaps \cite{Anderson2005NuPhA}. The
critical temperature profiles for these models are shown in
Fig.~\ref{fig:kappa-eta-And-APR}  for the HHJ EOS and in
Fig.~\ref{fig:kappa-eta-And-BSk} for the BSk21 EOS with right
vertical scales. The model `f' describes weaker proton
superconductivity based on the calculations in
Ref.~\cite{Amundsen1985NuPhA}. The corresponding panels in
Figs.~\ref{fig:kappa-eta-And-APR} and \ref{fig:kappa-eta-And-BSk}
are marked `weak SC'. Panels (b) and (d) in the same Figs., marked
`strong SC', show results for stronger proton superconductivity
model `e' that is fitted to the results of
Ref.~\cite{Elgaroy1996PhRvL77}, see Ref.~\cite{Anderson2005NuPhA}
for details. With the  same right vertical scale in each panel the
corresponding density-dependence of the parameter $A$  is shown.
Vertical dashed lines divide the regions of the superconductivity
of the first and second types, according to the criterion
$A>\sqrt{2}\pi$ ($\varkappa <1/\sqrt{2}$). Notice that the
critical temperature profiles for the same superconductivity
models are different for different EOSs because they actually
depend on the proton Fermi momentum $p_{\mathrm{F}p}$ which
differs in the HHJ and BSK21 EOSs at the same $n_B$.

The results for the shear viscosity $\eta_{e\mu}$ for the models
`f' and `e' are shown in the panes (a) and (b), respectively, in
Figs.~\ref{fig:kappa-eta-And-APR} and \ref{fig:kappa-eta-And-BSk}.
The thermal conductivity calculations are shown in panels (c) and
(d) of the same figures. All calculations employ  now $T=10^7$~K
in order to meet the zero-temperature approximation in a whole
density region shown (the low-temperature approximation can break
down near the walls of the critical density profile, where
$T_{\mathrm{c}p}$ is low). All the values can be scaled by $T^2$
for shear viscosity (by $T$ for thermal conductivity) provided the
condition $T<0.35T_{\mathrm{c}p}$ is fulfilled in the density
region of interest. As in Figs.~\ref{fig:kappa}--\ref{fig:eta},
solid lines in
Figs.~\ref{fig:kappa-eta-And-APR}--\ref{fig:kappa-eta-And-BSk}
show the results of the full calculations, while dashed and
dot-dashed lines are calculated in the London and Pippard limits,
respectively. Dotted lines represent the lepton transport
coefficients in normal matter. The results in
Figs.~\ref{fig:kappa-eta-And-APR}--\ref{fig:kappa-eta-And-BSk}
follow the same pattern as discussed above. The use of any of the
limiting expressions, London or Pippard, for the transverse
screening leads to an overestimation of the transport
coefficients. The London limit is appropriate in the case of
type-II superconductivity, while in the case of type-I
superconductivity, the London limit is inappropriate and the
Pippard limit can be a better approximation. In the intermediate
case full calculations should be used. As seen from
Figs.~\ref{fig:kappa-eta-And-APR}--\ref{fig:kappa-eta-And-BSk}, in
the real situation, both types of superconductivity can  be
simultaneously present in the NS cores
\cite{Glampedakis2011MNRAS}.

For comparison, in
Figs.~\ref{fig:kappa-eta-And-APR}--\ref{fig:kappa-eta-And-BSk} I
also show the neutron shear viscosity $\eta_n$ and thermal
conductivity $\kappa_n$ calculated following
Refs.~\cite{Shternin2013PhRvC,Shternin2017JPhCS}. In these Refs.,
the in-medium nucleon-nucleon interaction is treated in the
Brueckner-Hartree-Fock framework with the inclusion of the
effective three-body forces. The hatched strips in
Figs.~\ref{fig:kappa-eta-And-APR}--\ref{fig:kappa-eta-And-BSk}
show the results for normal (non-superconducting)
beta-equilibrated matter, and the widths of the strips illustrate
the uncertainty in calculations related to the different models of
the nuclear interactions as considered in
Ref.~\cite{Shternin2017JPhCS}. The lower boundaries correspond to
the nuclear interaction described by the Argonne v18 potential
with addition of the three-body forces in the phenomenological
Urbana~IX model. Upper boundaries correspond to the same
potential, but another  model for the three-body interaction based
on the meson-nucleon model of the nucleon interactions. More
details can be found in
Refs.~\cite{Baldo2014PhRvC,Shternin2017JPhCS}. Filled strips in
Figs.~\ref{fig:kappa-eta-And-APR}--\ref{fig:kappa-eta-And-BSk}
represent calculations of $\kappa_n$ and $\eta_n$ for the
proton-superconducting matter. As in the case of the lepton
transport coefficients, these results are obtained by neglecting
the collisions with protons (they are damped exponentially in the
considered limit). Then the neutron contribution to transport
coefficients is mediated by the neutron-neutron collisions only.
Notice, that the results for the neutron transport coefficients
shown in
Figs.~\ref{fig:kappa-eta-And-APR}--\ref{fig:kappa-eta-And-BSk}
include the corrections to variational solution
\cite{Shternin2013PhRvC,Shternin2017JPhCS}. Since $\kappa_n$ and
$\eta_n$ are calculated within specific models of the nucleon
interaction, the obtained results are not self-consistent with
EOSs used elsewhere in the present paper.  However, one expects
that the results shown here give plausible estimates for the
nucleon contribution (see Refs.~\cite{Shternin2013PhRvC} and
\cite{SchmittShternin2017arXiv} for more discussion).

Since neutron-proton collisions are damped in the superconducting
matter, respective $\kappa_n$ and $\eta_n$ values are larger than
those in normal matter (see, e.g.
\cite{Baiko2001A&A,ShterninYakovlev2008PhRvD}). However this
increase is much smaller than the increase in lepton transport
coefficients, which are additionally boosted by the change of the
screening behavior. According to
Figs.~\ref{fig:kappa-eta-And-APR}--\ref{fig:kappa-eta-And-BSk},
the relation between lepton and neutron transport coefficients
remains qualitatively the same as in the non-superfluid matter.
Namely, $\kappa_n>\kappa_{e\mu}$, while $\eta_n<\eta_{e\mu}$.
Remember (Sec.~\ref{sec:kin_form}) that the neutron transport
coefficients obey the standard Fermi-liquid behavior
$\kappa_n\propto T^{-1}$, $\eta_n\propto T^{-2}$ as do the lepton
transport coefficients in proton-superconducting matter.

All calculations above rely on the zero-temperature approximation
$T/\Delta\lesssim 0.2$. In this limit, it is enough to use the
expressions (\ref{eq:Pit_static})--(\ref{eq:J_T0})  for the proton
part of the transverse screening. Since this screening is static,
the lepton dynamical screening, which in the leading order is
proportional to $\omega$, see Eq.~(\ref{eq:Pit_norm}), was
neglected. In the Pippard limit this is possible when  $ \omega\ll
\pi\Delta/r$, where
$r=(p_{\mathrm{F}e}^2+p_{\mathrm{F}\mu}^2)/p_{\mathrm{F}p}^2\approx
1$. Since typically $\omega\sim T$, this is always justified in
our approximation. In the London limit, the similar comparison
requires $ \omega\sim T\ll 4/(3\pi) q_M v_{\mathrm{F}p}/r$. This
requirement becomes stronger with lowering density, and transforms
to  $T\ll 10^9$~K for  the BSk21 and HHJ EOSs at $n_B\approx 0.5
n_0$.

When t emperature starts to increase, the superfluid density of
protons $n_{sp}$ decreases and the Meissner momentum $q_M\propto
n_{sp}^{1/2}$ also decreases. The screening becomes
temperature-dependent. However, at low $q$ it is approximately
constant, moreover the change of the screening behavior from the
London one to the Pippard one occurs at the
temperature-independent value $q\sim\xi_0^{-1}$, where
$\xi_0\equiv \xi(T=0)$ \cite{Landau9eng}. Therefore,
qualitatively, the results of the above analysis hold if one takes
$A=\pi q_M(T) \xi_0$ (now $A$ is not related to $\varkappa$ which
is independent of $T$). At a given density, $A$ decreases with
increase of temperature making London limiting expressions more
appropriate. Transport coefficients start to decrease, and in the
leading order their temperature behavior is given by
Eqs.~(\ref{eq:kappa_emu_Meiss})--(\ref{eq:eta_emu_Meiss}),
providing temperature-dependent $q_M(T)$ is used in this case.
Such approach is possible until the dynamical part of the proton
polarization function and the lepton contribution
(\ref{eq:Pit_norm}) start to be important. Thus, at the
intermediate temperature the $\omega$ dependence of the transverse
polarization function needs to be taken into account, that
complicates the calculations
\cite{ShterninYakovlev2007PhRvD,ShterninYakovlev2008PhRvD}. In the
same region, the lepton-proton collisions start to be important
that additionally decrease the transport coefficients.  The
consideration of the lepton-proton collisions is less
straightforward since in the region of small momenta $q
v_{\mathrm{F}p}\sim\Delta$ the renormalization of the proton
current is necessary (e.g.,
\cite{KunduReddy2004PhRvC,Arseev2006PhyU}). In addition, because
of the gap in the proton spectrum, typical transferred energy is
of the order of $\Delta$ and the limiting approximation $q
v_{\mathrm{F}p}\gg\omega$ is not justified in the London limit.
Fortunately, due to the exponential suppression of the
lepton-proton collision frequencies, these effects need to be
taken into account relatively close to the critical temperature
where $\Delta(T)\sim T$. Then the approximation
$qv_{\mathrm{F}p}\gg \omega$ is valid since $\omega\sim\Delta\sim
T$. Clearly, the calculations of the transport coefficients in the
transition region $0.35 T_{\mathrm{c}p}\lesssim T\lesssim
T_{\mathrm{c}p}$ are more involved than in the simple
zero-temperature case. However, it seems sufficient in
applications to construct the smooth interpolation between the
results of the present paper at $T<0.35 T_{\mathrm{c}p}$ and the
normal matter results at $T>T_{\mathrm{c}p}$.

\section{Conclusions}\label{sec:conclusions}
I have calculated the electron and muon shear viscosity
$\eta_{e\mu}$ and thermal conductivity $\kappa_{e\mu}$ in the
proton-superconducting core of the NS based on the transport
theory of the Fermi systems. The present results are applicable at
the low temperatures, $T\lesssim 0.35 T_{\mathrm{c}p}$, and differ
from available calculations
\cite{ShterninYakovlev2007PhRvD,ShterninYakovlev2008PhRvD} by the
corrected account of the screening of the electromagnetic
interaction when the protons are in the  paired state.

The variational results for the thermal conductivity and for the
shear viscosity are obtained from
Eqs.~(\ref{eq:kappa_eta})--(\ref{eq:collfreq_system}) using the
appropriate collision frequencies. According to
Secs.~\ref{sec:collfreq_lead}--\ref{sec:collfreq_corr},
Eq.~(\ref{eq:kappa_lead}) with Eq.~(\ref{eq:fit0}) can be used for
thermal conductivity calculations. For the shear viscosity, it is
enough to use the leading-order contributions in $q$ in
Eqs.~(\ref{nueta_ci})--(\ref{nueta_ci_tlfull}). The explicit
expressions for these contributions are given by
Eqs.~(\ref{eq:nu_eta_ci_t_lead})--(\ref{eq:nu_eta_ci_tl_lead}).
Finally, the simplest variational solution  works well for
$\eta_{e\mu}$, while for $\kappa_{e\mu}$ additional factor
$C^\kappa=1.2$ should be used in Eq.~(\ref{eq:kappa_eta}) to
correct the variational result.

The main conclusions of the present study are as follows:
\renewcommand{\labelenumi}{(\roman{enumi})}
\begin{enumerate}
\item In the superconducting NS cores, lepton transport
coefficients obey the standard Fermi-liquid temperature dependence
$\kappa_{e\mu}\propto T^{-1}$, $\eta_{e\mu}\propto T^{-2}$ in
contrast to the situation in normal NS cores where
$\kappa_{e\mu}\approx \mathrm{const}(T)$, $\eta_{e\mu}\propto
T^{-5/3}$. This is a consequence of the static regime of the
screening of electromagnetic interactions. The screening in the
transverse channel is dominated by the proton contribution.
\item At a given density, $\kappa_{e\mu}$ and $\eta_{e\mu}$
increase with increase of the proton critical temperature
$T_{\mathrm{c}p}$ (increase of the gap $\Delta$). At large
densities, where the typical transferred momentum $q$ is large so
that the Pippard limit for the transverse screening is applicable,
one finds $\kappa_{e\mu}\propto \Delta$ and $\eta_{e\mu}\propto
\Delta^{1/3}$. In the opposite limit of low densities, where the
transverse electromagnetic interaction is screened by the Meissner
momentum (the London limit), $\kappa_{e\mu}$ and $\eta_{e\mu}$ are
independent of $\Delta$.
\item In general situation relevant for the NS cores, the whole
range of momentum transfer is important, both limiting expressions
overestimate transport coefficients, and the complete results
developed here shall be used. However, in case of the
superconductivity of the second kind it is  enough to take the
London limit for transverse screening and use corresponding
limiting expressions for the transverse part of the collision
frequencies. In case of the type-I superconductivity, the Pippard
limit is more appropriate, although it is recommended to rely on
the complete result in this limit.
\item Both limiting expressions can be used to estimate  the
transport coefficients from above. In this respect, the expression
in London limit is more interesting since it gives the
gap-independent boundary.
\item As in the case of the normal matter
\cite{Shternin2013PhRvC,SchmittShternin2017arXiv}, leptons give
dominant contribution to the shear viscosity while neutrons
dominate the thermal conductivity.
\end{enumerate}

Since the consideration of the lepton transport coefficients
presented here does not rely of the specific EOS properties, the
conclusions (i)--(iv) are universal in a sense that they are valid
for any npe$\mu$ EOS of the dense matter in NS cores. In contrast,
the conclusion (v) is model-dependent and need to be taken with
caution. The neutron transport coefficients shown in
Figs.~\ref{fig:kappa-eta-And-APR}--\ref{fig:kappa-eta-And-BSk} are
calculated for different nucleon interactions, but within the
single many-body approach, namely the Brueckner-Hartree-Fock
scheme. Thus it is in principle not excluded that the conclusion
(v) can fail for a specific microscopic model that produces
significantly different values of $\kappa_n$ and $\eta_n$ than
used here. Ideally,  the neutron transport coefficients need to be
calculated from the same microscopic model of the nucleon
interaction as the EOS which is not a straightforward task. The
detailed discussion of the neutron transport coefficients is
outside the scope of the present paper, see
Refs.~\cite{Shternin2013PhRvC,SchmittShternin2017arXiv} for more
details.

The results obtained in the present paper can be improved by
considering the finite temperature effects in order to study in
detail the transition from the superconducting to the normal
matter. This can be done following the same lines as described
here, but considering the full temperature-dependent polarization
functions. In this case, however, one needs to account for the
dynamical part of the screening making the interaction
$\omega$-dependent. In this case the integration over $\omega$ in
Eqs.~(\ref{eq:nu_kappa})--(\ref{eq:nu_eta1}) cannot be performed
analytically. Additional care must be taken when lepton collisions
with the protonic excitations are considered. Anyway, it seems
enough to interpolate through the transition region for practical
applications, however the detailed investigation of the transition
region remains to future studies.

In this paper I used polarization functions calculated in pure BCS
framework, neglecting Fermi-liquid effects. In the Pippard limit
this is a good approximation \cite{Gusakov2010PhRvC}. However, the
static screening in the London limit is affected
\cite{Gusakov2010PhRvC}. The consideration of these effects
requires separate study. Moreover, it was proposed that the
coupling between neutrons and protons in NS cores induces the
effective electron-neutron interaction \cite{Bertoni2015PhRvC}
that modifies the screening properties of matter
\cite{Stetina2018PhRvC} and can affect the lepton collision
frequencies. The effect of this interaction on the transport
coefficients can be expected both in the normal and
superconducting matter and is under investigation
\cite{Stetina2018PhRvC}.

In the inner cores of NSs,  hyperons can also appear
\cite{HPY2007Book}. If they are normal (unpaired), the results of
the present paper for lepton transport coefficients can be easily
generalized by treating charged hyperons as passive scatterers. It
is thought, however, that hyperons like the protons can be paired
in the $^1S_0$ channel (since their number density is low), see
e.g. review in Ref.~\cite{Sedrakian2018arXiv}. In this case, their
contribution to the transverse plasma screening should be
considered in the same way as the proton one, although the
situation will be more cumbersome since more than one gap is
involved.

\acknowledgments The author is grateful to N.~Chamel, M.~E.
Gusakov, and D.~G. Yakovlev for discussions. The work was
supported by the Foundation for the Advancement of Theoretical
Physics and Mathematics ``BASIS'' and the Russian Foundation for
Basic Research, grant 16-32-00507 mol$\_$a.
\bibliographystyle{apsrev}
\def\apjl{Astrophys. J. Lett.}
\def\apjs{Astroph. J. Suppl. Ser.}
\def\mnras{Mon. Not. R. Astron. Soc.}
\def\aap{Astron. Astrophys.}
\def\apss{Astroph. Space Sci.}
\def\pla{Phys. Lett.  A}
\def\ssr{Space Sci. Rev.}
\def\araa{Ann. Rev. Astron. Astrophys.}
\def\aj{Astron. J.}
\def\jphys{J. Phys.}
\def\npa{Nucl. Phys. A}
\def\npb{Nucl. Phys.  B}
\def\ijmpe{Int. J. Mod. Phys. E}
\def\ijmpd{Int. J. Mod. Phys. D}
\def\ijmpa{Int. J. Mod. Phys. A}

\begin{appendix}*
\section{Fitting expressions for integrals}
The transverse integrals (\ref{eq:In}) can be normalized as
\begin{equation}
I_n=q_M^{3-n} \tilde{I}_n(x_t,A),
\end{equation}
where $x_t=q_M/q_m$ and
\begin{equation}
\tilde{I}_n(x_t,A)=\int\limits_0^{1/x_t} \frac{x^n d\,x}{\left(x^2+J(A x)\right)^2}.
\end{equation}
The integrals $\tilde{I}_n(x_t,A)$ were calculated on the dense grid and fitted by analytical expressions that take into account the correct asymptotic behavior in the limiting cases.

For $n=0$ the result is
\begin{equation}\label{eq:fit0}
\frac{p_1 A^{1.5}+p_2A^2+\pi/4-x_t^3/3+p_3 x_t^4+A \left(p_4-p_5 x_t^3\right)}{1+p_6A},
\end{equation}
where $p_1=0.0119$, $p_2=0.0063$, $p_3=0.202$, $p_4=0.108$,
$p_5=0.454$, and $p_6=0.14$. Fit rms error is 0.4\% and the
maximal fitting error is 3\% at $A=5$ and $x_t=0.76$.

Similarly, for $n=2$.
\begin{eqnarray}
&&(1+p_6A)^{-1}\left[p_1 A^{4/3} +\frac{\pi}{4}-x_t-p_2A x_t+p_3 x_t^2\right.\nonumber\\
&&\left.+A^{2/3} (p_4+p_5 x_t)+A^{1/3} \left(p_7-p_8 x_t+p_9 x_t^2\right)\right], \nonumber\\
\label{eq:fit2}
\end{eqnarray}
where $p_1=0.094$, $p_2=0.234$, $p_3=0.356$, $p_4=0.1859$,
$p_5=0.0496$, $p_6=0.227$, $p_7=-0.0537$, $p_8=-0.2345$, and
$p_9=0.239$. Fit rms error is $\sim1$\% and the maximal fitting
error is 7\% at $A=0.5$ and $x_t=0.14$.

The leading-order contributions to the collision frequencies
relevant for the shear viscosity problem
(\ref{nueta_ci_tfull})--(\ref{nueta_ci_tlfull}) can be given as
\begin{equation}
\nu^{\eta,t}_{ci}=\frac{2\pi \alpha_f^2 T^2 p_{\mathrm{F}i}^2 }{m_c^* p_{\mathrm{F}c}}I^t_2,\label{eq:nu_eta_ci_t_lead}
\end{equation}
\begin{equation}
\nu^{\eta,l}_{ci}=\frac{2\pi \alpha_f^2 T^2 m_c^* m_{i}^{*2}}{p_{\mathrm{F}c}^3q_\mathrm{TF}}\left(\mathrm{arctan}\frac{q_m}{q_\mathrm{TF}}+\frac{q_\mathrm{TF}q_m}{q_\mathrm{TF}^2+q_m^2}\right), \label{eq:nu_eta_ci_l_lead}
\end{equation}
\begin{equation}
{\nu'}^{\eta}_{ci}=\frac{4\pi \alpha_f^2 T^2 m_i^*p_{\mathrm{F}i}}{p_{\mathrm{F}c}^2 q_\mathrm{TF}}\mathrm{arctan}\frac{q_m}{q_\mathrm{TF}}.\label{eq:nu_eta_ci_tl_lead}
\end{equation}

\end{appendix}
\end{document}